%% file: main.tex
\documentclass[conference]{IEEEtran}
\usepackage{cite}
\usepackage{amsmath,amssymb,amsfonts}
\usepackage{graphicx}
\usepackage{textcomp}
\usepackage[hyphens]{url}
\usepackage{fancyhdr}
\usepackage[bookmarks=true,breaklinks=true,letterpaper=true,colorlinks,citecolor=blue,linkcolor=blue,urlcolor=blue]{hyperref}

\usepackage[bookmarks=true,breaklinks=true,letterpaper=true,colorlinks,citecolor=blue,linkcolor=blue,urlcolor=blue]{hyperref}
\usepackage{graphicx}
\usepackage{subcaption}
\usepackage{listings}
\usepackage[table]{xcolor}
\usepackage{textcomp}
\usepackage{boldline}
\usepackage{float}
\usepackage{xspace}
\usepackage{booktabs, makecell, multirow, tabularx}
\usepackage{todonotes} 
\usepackage{wrapfig,lipsum}
\usepackage{verbatim}
\usepackage{graphicx}
\usepackage{listings}
\usepackage{tikz}
\usepackage{multicol}
\usepackage{wrapfig}
\usepackage{kantlipsum}
\usepackage{boldline}
\usepackage{bbding,captdef}
\usepackage{url}
\usepackage{algorithm}
\usepackage{soul}
\usepackage{algpseudocode}
\newcommand{\ifmap}{\textit{ifmap}\xspace}
\newcommand{\ofmap}{\textit{ofmap}\xspace}
\newcommand{\ifmaps}{\textit{ifmaps}\xspace}
\newcommand{\ofmaps}{\textit{ofmaps}\xspace}
\newcommand{\fmap}{\textit{fmap}\xspace}

\definecolor{dkgreen}{rgb}{0,0.6,0}
\definecolor{gray}{rgb}{0.5,0.5,0.5}
\definecolor{mauve}{rgb}{0.58,0,0.82}

\newcommand*\circlenew[1]{\tikz[baseline=(char.base)]{
            \node[shape=circle,draw,inner sep=0.3pt] (char) {#1};}}      

\usepackage{tikz}
\newcommand*\emptycirc[1][1ex]{\tikz\draw (0,0) circle (#1);} 

\newcommand*\fullcirc[1][1ex]{\tikz\fill (0,0) circle (#1);} 
\algnewcommand{\Initialize}[1]{%
  \State \textbf{Initialize:}
  \Statex \hspace*{\algorithmicindent}\parbox[t]{.8\linewidth}{\raggedright #1} }   

\newcommand{\name}{{\em NeuroPlug}\xspace}
\newcommand{\fname}{{\em NeuroPlug}\xspace}
\def\BibTeX{{\rm B\kern-.05em{\sc i\kern-.025em b}\kern-.08em
    T\kern-.1667em\lower.7ex\hbox{E}\kern-.125emX}}

\newenvironment{stbox}[1]{
  \begin{tcolorbox}[colback=gray!5!white,colframe=gray!75!black,title=#1,
    left=0.2pt,right=0.2pt,top=0.2pt,bottom=0.2pt] 
}{
  \end{tcolorbox}
}

\usepackage{cancel}
\usepackage{tikz}                   
\usetikzlibrary{shadows}
\usepackage{fancybox} 
\usepackage{tcolorbox}
\usepackage{array}
\tcbuselibrary{skins}
\definecolor{royalblue}{RGB}{65,105,225}

\fancypagestyle{firstpage}{
  \fancyhf{}
  
  \fancyfoot[C]{\thepage}
}

\newcommand{\blue}[1]{\textcolor{black}{#1}}

\title{NeuroPlug: Plugging Side-Channel Leaks in NPUs using Space Filling Curves}

\author{
\IEEEauthorblockN{Nivedita Shrivastava}
\IEEEauthorblockA{Dept. of Electrical Engineering\\
Indian Institute of Technology Delhi\\
New Delhi, India\\
nivedita.shrivastava@ee.iitd.ac.in}
\and
\IEEEauthorblockN{Smruti R. Sarangi}
\IEEEauthorblockA{Dept. of Electrical Engineering\\
Indian Institute of Technology Delhi\\
New Delhi, India\\
srsarangi@cse.iitd.ac.in}}

\begin{document}
\maketitle
\thispagestyle{firstpage}
\pagestyle{plain}


\input{abstract}
\input{Introduction_ne}
\input{background}
\input{threatModel}
\input{CMs}
\input{Motivation}
\input{theory}
\input{design}
\input{evaluation}
\input{securityEvaluation}
\input{relatedWork}
\input{conclusion}


\bibliographystyle{IEEEtranS}
\bibliography{ref}

\end{document}

%% file: abstract.tex
\begin{abstract}
Securing deep neural networks (DNNs) from side-channel attacks is an important
problem as of today, given the substantial investment of time and
resources in acquiring the raw data and training complex models.
All published countermeasures (CMs) add noise $N$ to a signal $X$ (parameter of interest such
as the \blue{net memory traffic} that is leaked). 
The adversary observes $X+N$; we shall show that
it is easy to filter this noise out using targeted
measurements, statistical analyses and different kinds of reasonably-assumed side information.
We present a novel CM \fname that is immune to these attack methodologies mainly because we use a different formulation
$CX+N$. We introduce a multiplicative variable $C$ that \blue{naturally} arises from
feature map compression; \blue{it} plays a key role in
obfuscating the parameters of interest. 
Our approach is based on mapping all the computations to a 1-D space filling curve and then
performing a sequence of tiling, compression and binning-based obfuscation operations. We follow up
with proposing a theoretical framework based on Mellin transforms that allows us to accurately
\blue{quantify} the size of the search space as a function of the noise we add and the side information
that an adversary possesses. The security
guarantees provided by \fname are validated using a battery of statistical and information theory-based tests. We also
demonstrate a substantial performance enhancement of 15\% compared to the closest competing work.
\end{abstract}

%% file: Introduction_ne.tex
\section{Introduction}

Deep neural network (DNN) 
models embody a significant amount of intellectual property (IP).
Creating such models requires a massive amount of investment in terms of financial resources and human effort:
data collection via mechanical turks, model design and rigorous training~\cite{securator,scedl}.
Attackers thus have a very strong incentive to steal the model parameters associated with these
DNN models such as the nature/number of layers, details of weights, etc.~\cite{securator,scedl}. 
\blue{Such security breaches can potentially compromise the privacy, security and integrity of large autonomous systems.}
For example, an attacker can make a small human-invisible change to lane markings or traffic lights, which are sufficiently powerful to confuse lane detection and traffic light sensing models, respectively~\cite{trafficSign1,trafficSign2,trafficLight}.
Possessing even partial knowledge about the inner workings of a DNN model significantly enhances the feasibility of orchestrating successful attacks --
adversarial attacks~\cite{adversarial}, membership
inference attacks~\cite{membership} and bit-flip attacks~\cite{deephammer}.  

For the last 5 years, attackers and defenders have been locked
in an arms race to design more potent attacks and countermeasures (CMs),
respectively. The attacker relies on a memory/timing based side channel that leaks a signal $X$, 
which carries a lot of information about the aforementioned model parameters. \blue{$X$
could be the number of off-chip bytes transferred ({\em volume}), computational {\em time}, 
the number of bytes read between a write$\rightarrow$read
to the same memory address (read-write {\em distance}), the number of times a byte is read ({\em count}), or any
combination thereof. }
Existing countermeasures (CMs)
present $Y=X+N$ to the attacker, where $N$ is the \blue{added} noise. 
State-of-the-art attacks~\cite{reverse,huffduff,cachetelepathy,deepsniffer} look at a subset of these four variables
-- count, distance, time and volume (CDTV) -- and try to reduce the search space of $X$ by collecting multiple 
measurements and
using specially crafted inputs.  On the other hand, CMs use a combination of shuffling data, adding dummy accesses
and on-chip buffering to hide the noise parameter. 


\begin{table*}[!htb]
 \footnotesize
    \centering
 \begin{tcolorbox}[enhanced, width=0.958\linewidth, boxsep=0pt, left=0pt,right=0pt,top=0pt,bottom=0pt,
    colback=white, colframe=black, arc=0mm, boxrule=0pt,
    drop shadow={shadow xshift=100mm, shadow yshift=100mm}]
   
 \rowcolors{2}{royalblue!20}{royalblue!5}
   
    \begin{tabular}{|l|cccc|ccc|c|c|c| }
    \hline
    \rowcolor{gray!10}
     \textbf{Countermeasure} &  \multicolumn{4}{c|}{\textbf{Protects}} & \multicolumn{3}{c|}{\textbf{Security Metric}} & \textbf{Perf.} & \textbf{Counterattacks} & \textbf{Leakage}  \\
     \cline{2-8}
    \rowcolor{gray!10}
        & \textbf{Count} & \textbf{Distance} & \textbf{Time} & \textbf{Volume} & \textbf{SrS.} & \textbf{Accu.} & \textbf{Theo.} & \textbf{metric}& &  \\
    \hline

   Liu et al.~\cite{mitigating}   &  \emptycirc[1ex] & \emptycirc[1ex] & \emptycirc[1ex] & \fullcirc[1ex] & \fullcirc[1ex] & \emptycirc[1ex]   & \emptycirc[1ex]  & \fullcirc[1ex] & Statistical ($\S$~\ref{sec:SS}) & RW Dependencies\\  

   NPUFort~\cite{npufort} &  \emptycirc[1ex] & \emptycirc[1ex] & \fullcirc[1ex]. & \emptycirc[1ex] & \emptycirc[1ex] & \emptycirc[1ex]   &  \emptycirc[1ex] & \fullcirc[1ex] & Side-infor. ($\S$~\ref{sec:SI}) & Mem. access pattern\\
       
   NeuroObfus.~\cite{neurobfuscator}  & \fullcirc[1ex] & \emptycirc[1ex] & \emptycirc[1ex] & \fullcirc[1ex] & \emptycirc[1ex]   & \fullcirc[1ex]  & \emptycirc[1ex]  &  \fullcirc[1ex] & Statistical ($\S$~\ref{sec:KK}) & Mean noise\\

   Zuo et al.~\cite{sealing} & \emptycirc[1ex] & \emptycirc[1ex] & \fullcirc[1ex]. & \emptycirc[1ex] & \emptycirc[1ex] & \fullcirc[1ex]  &  \emptycirc[1ex] & \fullcirc[1ex] & Side-infor. ($\S$~\ref{sec:SI}) & Mem. access pattern  \\
           
  ObfuNAS~\cite{obfunas} & \fullcirc[1ex] & \emptycirc[1ex] & \emptycirc[1ex] & \fullcirc[1ex] & \emptycirc[1ex] & \fullcirc[1ex]  &  \emptycirc[1ex] & \fullcirc[1ex]  & Statistical ($\S$~\ref{sec:KK}) & Mean noise\\

  DNNCloak~\cite{dnncloak} & \fullcirc[1ex] & \fullcirc[1ex] & \emptycirc[1ex] & \fullcirc[1ex] & \emptycirc[1ex] & \fullcirc[1ex]  &  \emptycirc[1ex]  & \fullcirc[1ex] & Side-infor. ($\S$~\ref{sec:SI}) & Values of RWs \\
 \rowcolor{royalblue!20}
   & &  & &  &  &   &   & & Kerckhoff ($\S$~\ref{sec:KK}) & Buffer size\\
\hline 
\rowcolor{gray!10}
       \textbf{\fname}  & \fullcirc[1ex] & \fullcirc[1ex] & \fullcirc[1ex] & \fullcirc[1ex] & \fullcirc[1ex] & \fullcirc[1ex]  & \fullcirc[1ex] & \fullcirc[1ex] & - & - \\
       \hline
       
    \end{tabular}
    \end{tcolorbox}
    \caption{ \fname offers comprehensive protection and undergoes thorough validation using various security metrics. \fullcirc[1ex] property is satisfied and \emptycirc[1ex] means it is not. \textbf{SrS.} is the search space, \textbf{Acc.} is model re-training accuracy or how accurately the layer parameters are estimated. \textbf{Theo.}  refers to information-theory based security metrics.}
    \label{tab:countermeasures}
\end{table*}

\noindent \textbf{Shortcomings of Current CMs} 
The most relevant, state-of-the-art CMs are shown in Table \ref{tab:countermeasures} (details in
Section~\ref{sec:back}). 
Given that $X$ is a constant for a given layer \blue{as the model parameters will remain the same}, repeated
measurements with possibly different inputs and possibly
across multiple devices shall yield the distribution of $N$. If the mean of $N$ is 0 or the minimum is 0, then the
noise gets effectively filtered out ({\em statistical (SS) attack})(\cite{SS1,mitigating}). 
Assume that this is not the case, then any measurable statistical metric
\blue{is a non-zero} constant $N_c$, which is typically a hardwired value~\cite{dnncloak}.  Any such
hardwired value \blue{can be leaked by a malicious insider}~\cite{KK1,KK2,KK3,KK4}. \blue{It also
violates the Kerckhoff's principle~\cite{KP}, which clearly
states that security cannot be derived from hiding aspects of the design.}

Such {\em Kerckhoff (KK) attacks} can yield all the constants that characterize
the parameters of the noise distribution and this leads to an estimate of $X$. Now, assume that despite
statistical and Kerckhoff attacks, our search space is still large. We can then glean some knowledge from
sister neural networks (belonging to the same genre) and try to make educated guesses. For instance,
layer sizes are mostly multiples of squares(see\cite{reverse,dnncloak}. Such numbers are formally known as
non-square-free (NSQF) numbers~\cite{squareful}. With other
reasonable assumptions regarding layer skewness, our search space can reduce drastically. Let us refer
to this as a side-information (SI) attack.

Using these three attacks -- SS, KK and SI -- we were able to break all the CMs mentioned in
Table~\ref{tab:countermeasures}. Breaking these approaches required a maximum of 60,000 runs.
If the real fabricated NPUs were available (7nm technology), we estimate the process to take
a maximum of 8-10 hours. Using layer-based obfuscation approaches (dynamically fuse/split layers)
\cite{neurobfuscator} did not prove to be very useful other than enlarging the search space moderately,
primarily because it does not fundamentally obfuscate the CDTV metrics.

\noindent \textbf{Our Insights}
We introduce a new degree of freedom by augmenting
the $X+N$ formulation with a multiplicative random variable $C$.
The new formulation becomes an affine transform of the form \fbox{$CX+N$}. 
We shall show in Section~\ref{sec:eval} that for our proposed
design \fname, this formulation increases the search space by roughly $10^{70}$ times
after filtering out as much of the
noise as we can (using the SS, KK and SI attacks). This is the bedrock of our design. 

\noindent \textbf{Our Solution: \fname}
This factor $C$ \blue{naturally
arises} from data compression in our design and \blue{can further be augmented by adding custom noise}.
Insofar as the adversary is concerned, all the CDTV metrics are affected to different extents because of compression --
the effect is equivalent to reading and writing $CX$ bytes.
To enable data compression as an obfuscatory tool, we need
to represent a computation in a DNN differently. \blue{This choice segues into the novel notion of
using 1-D space filling curves (SFCs)~\cite{SFC} to
map a complex 3-D space of DNN computations to an equivalent 1-D space}.
This 1-D curve conceptually snakes through all
the computation layers of a DNN (possibly with skip connections) and
allows us to create contiguous chunks of bytes (tiles), \underline{compress them} and {\em bin} them.

A {\em tile} can be split across bins and we can also leave space within a bin empty (contribution to the $N$ in $CX+N$).
Now, $C$ is tied to the plaintext. Many may argue that given that the plaintext is known, it is possible to estimate
$C$, at least in the first few layers.  In the later layers due to both aleatoric (inherent) and epistemic (lack of
knowledge of model parameters) uncertainty, estimating $C$ is difficult~\cite{surveyUncertainty}. Hence, in the first few layers, we need to
\blue{enhance the uncertainty in the noise $N$ and also add some random text in the data to obfuscate $C$}. Plaintext-aided
encryption is known to work well with such safeguards~\cite{chaos1,chaos2,chaos3}. 
Now, to not run afoul of the SS and KK attacks, we make these dummy text snippets and all the parameters that determine
the nature of the noise distribution $N$ a part of the {\em key} (the encrypted DNN model in this case). {This is allowed}. \blue{This ensures that even if the key is compromised, the effects will be limited to systems that
use the key. The key can always be changed}
(refer to the paper on secure leasing~\cite{securelease}). 

\begin{figure}[!htb]
    \centering
    \includegraphics[width=\linewidth]{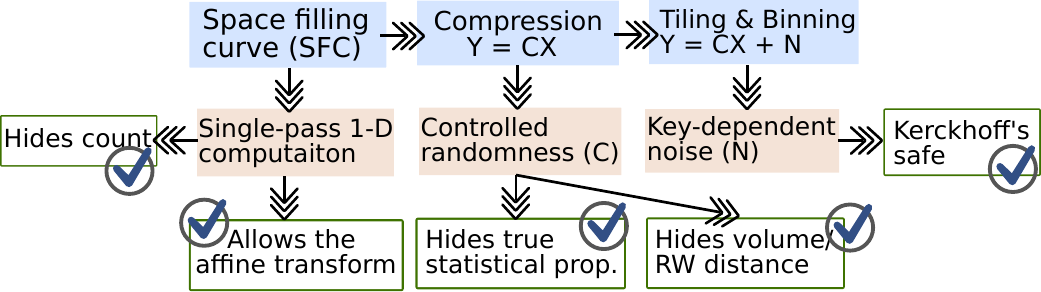}
    \caption{The overall scheme}
    \label{fig:overall}
\end{figure}

Figure~\ref{fig:overall} shows the outline of our scheme. The SFC and the affine transform ($CX+N$) are the central
themes, which lead to an architecture based on tiling, binning, compression and noise addition. Regardless of the
side-information available, we show that introducing the extra degree of freedom ($C$) has a dramatic
effect on the search space, \blue{whose size is computed using our novel 
Mellin transform~\cite{mellin} based framework. Given that we can modulate $C$ by adding random noise, the
size of the search space can be increased (at the cost of performance).  This may be required in the future if the
model owner desires greater levels of protection or the HW becomes faster}. To the best
of our knowledge, such a systematic and mathematically grounded approach of search space estimation is novel.
\blue{Note that the size of the search space yielded by our analysis
is a function of the side information possessed by the adversary. The adversary is limited to only memory
and timing-based side channels.}

The specific \textbf{list of contributions in this paper} is as follows:  
\circlenew{1} A comprehensive experimental
study of the information leaked regarding input/output feature dimensions by sister NNs.
\circlenew{2} A thorough analysis of the limitations in existing state-of-the-art security solutions in the light of recent threats -- SS, KK and SI attacks.
\circlenew{3} A formal theoretical framework to quantify the ``smart'' search space in the system  given a bound on the information leaked to the adversary.
\circlenew{4} A novel approach to
enhance the security of NN accelerators hiding the leaked signatures using an SFC-based execution pattern.  
\circlenew{5} A detailed theoretical and performance analysis of
\fname, which shows a $15\%$ performance improvement over the closest competing work (\textit{DNNCloak}). 
\circlenew{6} Two case studies of state-of-the-art attacks on sparse and dense accelerators (resp.), 
a rigorous evaluation based on computing the size of the search space for a given amount of added noise (and side
information), \blue{a performance-security trade-off analysis}, and
thorough statistical and information theory-based tests. 

\noindent {$\S$\ref{sec:back} provides the background and outlines the threat model. $\S$\ref{sec:motivation} provides the motivation.} $\S$\ref{sec:theory} provides a theoretical description of the proposed scheme. $\S$\ref{sec:hw}
presents the proposed design, $\S$\ref{sec:eval} and $\S$\ref{sec:security} present the performance and security analyses, respectively. $\S$\ref{sec:RW} presents the related work. We finally conclude in
$\S$\ref{sec:conclusion}. 

%% file: background.tex
\section{Background and Threat Model} \label{sec:back}
\subsection{Convolution Neural Networks (CNNs)}
Convolution serves as the core component of a DNN. 
To process a layer, we sequentially
traverse the input and output feature maps (referred to \ifmaps and \ofmaps, respectively), and apply the convolution
operation. The $K$ \ofmaps, each of size ($P \times Q$) are generated by convolving a filter of size ($R \times S$) with $C$ \ifmaps.
Each \ifmap's size is $ H \times W$. 
For the sake of explanation, we assume that the \ofmap and \ifmap have the same dimensions ($P=H; Q=W$) (refer to Table~\ref{tab:symbol}).

Due to the limited on-chip storage, the elements in an \ifmap, an \ofmap, and a filter are often grouped into {\em
tiles}~\cite{timeloop} as shown in Listing~\ref{lst:conv}. 


\begin{lstlisting} [caption={A basic convolution operation},captionpos=b,label=lst:conv]
// Off-chip 
for(k$_o$=0; k$_o$<K; k$_o$+=T$_k$)  // K: #output fmaps
 for(c$_o$=0; c$_o$<C; c$_o$+=T$_c$)  // C: #input fmaps
  for(h$_o$=0; h$_o$<H ; h$_o$+=T$_h$)  // H: #rows in a fmap   
   for(w$_o$=0; w$_o$<W ; w$_o$+=T$_w$)   // W: #cols in a fmap
// On-chip 
    for(k=k$_o$; k<k$_o$+T$_k$ ; k++)  // Tile level processing 
     for(c=c$_o$; c<c$_o$+T$_c$ ; c++)     
      for(h=h$_o$; h<h$_o$+T$_h$ ; h++)    
       for(w=w$_o$; w<w$_o$+T$_w$ ; w++)   
        for(r=0; r<R ; r++)  // R: #rows in a filter
         for(s=0; s<S ; s++) // S: #cols in a filter
           ofmap[k][h][w]+=ifmap[c][h+r][w+s] * weights[k][c][r][s];

\end{lstlisting}

\begin{table}[!htb]
    \centering
    \footnotesize
    \begin{tabular}{|p{.5cm}|l|p{.6cm}|l|}
    \hline
    \rowcolor{black!10}
    \textbf{Sym.} & \textbf{Notation} & \textbf{Sym.} & \textbf{Notation} \\
    \hline
       K & \#\ofmaps & T$_k$ & Tiling factor for \# \ofmaps \\
       C & \#channels & T$_c$ & Tiling factor for \# \ifmaps\\
       H/W & \#rows/cols in an \ifmap & T$_h$/$T_w$ & Tiling factor for rows/cols\\
       R/S & \#rows/cols in a filter	&  P/Q & \#rows/cols in a filter\\
       $\beta$ & Compression ratio & $\alpha$ & Additive noise\\
    \hline
    \end{tabular}
 \caption{Glossary of the symbols}
\label{tab:symbol}    
\vspace{-5mm}
\end{table}

%% file: threatModel.tex
\subsection{System Design and Threat Model}
\label{sec:threat}
\begin{figure}[!htb]
    \centering
    \includegraphics[width=0.75\linewidth]{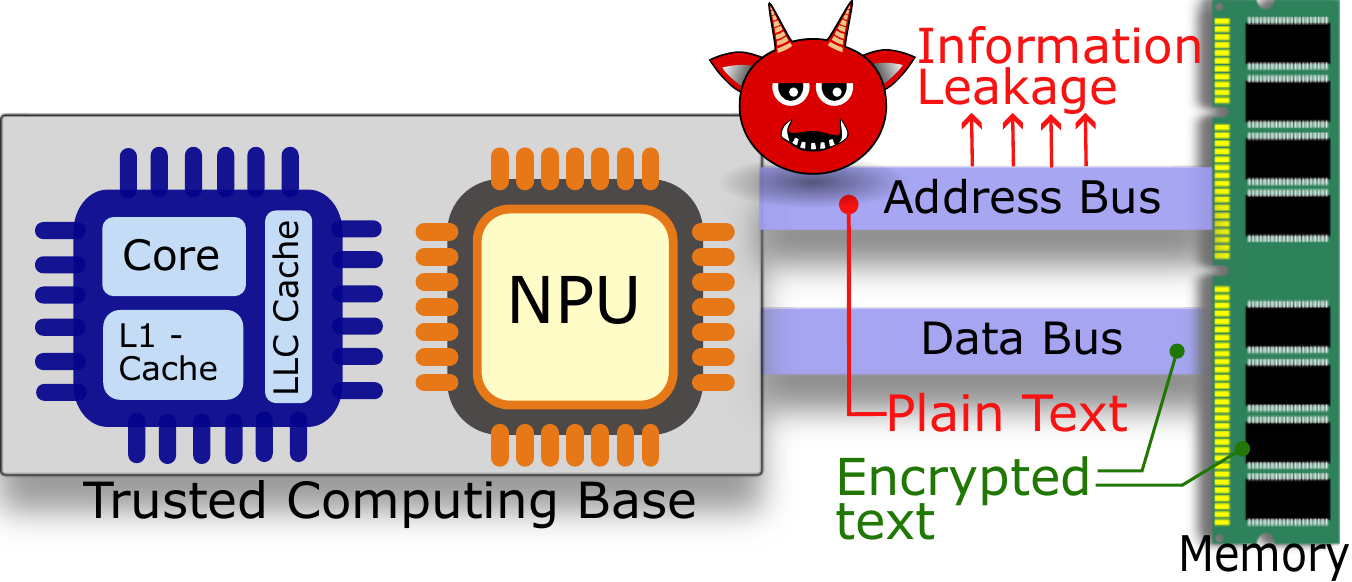}
    \caption{The system design and the threat model (similar to \cite{reverse,huffduff,neurobfuscator}). An adversary snoops the address and the data buses to infer the memory access patterns.}
    \label{fig:threat_model}
\end{figure}

Our threat model reflects a scenario where a DNN model is performing inference tasks locally on an 
NPU using a
pre-trained model. We present the high-level system design and threat model in Figure
\ref{fig:threat_model} (similar to previous attacks and CMs~\cite{reverse,huffduff,neurobfuscator}). 

We assume that the system consists of a central processing unit (CPU), NPU and caches. The CPU sends commands to
the NPU through a {\em secure} channel. Like prior work, 
all transfer of data between the CPU and NPU takes place
via main memory. The data in the main memory is \blue{encrypted and tamper-proof} (using~\cite{securator}).
We assume the existence of a trusted execution environment (TEE) like
Intel SGX~\cite{sgx}, which is used to download the secure model and get the decryption key via 
a trusted third-party attestation based mechanism (refer to~\cite{attestation}). It can also periodically renew
the model and the key using secure leasing~\cite{securelease}.
The NPU, CPU, caches and the TEE are all within the trusted computing base (standard design
decision:\cite{guardnn,tnpu,securator}). 

Potential adversaries might include the operating system, the hypervisor, a malevolent
application, or somebody having physical access to the main memory or memory bus. 

\noindent \textbf{Attacker's Aim} The primary objective of an adversary is to reverse-engineer the architecture of the
DNN model in order to facilitate subsequent attacks, including adversarial attacks and membership inference. She can
also aim to retrain a comparable model with enhanced or similar accuracy.   

\noindent \textbf{Attacker's Capabilities and Limitations} 
We assume a strong attacker, who can target the system memory and memory
interface. She can craft specialized inputs and feed them to the accelerator. \blue{The adversary possesses the ability to
read the memory addresses that are sent as plaintext
on the memory bus and identify the type of memory access instruction (read or write). She
can also monitor the transfer of encrypted data to/from DRAM memory.
She however cannot tamper with the addresses or the encrypted data, or mount replay attacks (this can be ensured via 
Securator~\cite{securator}).} She
can also use a high-resolution timer to track the timing between memory accesses. 
Then, she can mount SS, KK and SI attacks. This means that a model can be run any
number of times with the same or different inputs and it can be run on as many devices as needed (to thwart
CMs that use PUFs\cite{puf}). The attacker has full access to all hardwired parameters and the design of the NPU. 
Finally, the attacker has access to all contemporary DNN models and has learnt whatever it could from them.
This information can be leveraged to mount SI attacks (we will quantify this in Section~\ref{sec:theory}).

\blue{We do not consider power, thermal and EM-based attacks. Cache based attacks are not relevant
here because the CPU's cache hierarchy is not used.}

%% file: CMs.tex
\subsection{State-of-the-art Attacks and Countermeasures (CMs)}

\blue{An attack aims to find the parameters for each CNN layer: size, width, height, number of
\ifmaps (input channels) and \ofmaps (output channels). 
Reading a fully-computed \ofmap signifies the termination of the preceding layer that
produced it. This 
helps the adversary identify layer boundaries. 
There is a RAW dependency here (across layers) that is characterized by the write-to-read \textit{distance}
(bytes accessed between a write and the corresponding read). This is indicative of the layer size. 
Next, note that the same \ifmap is used to generate different \ofmaps using a
plurality of filters. This means that the same datum is accessed several times (non-zero \emph{count}):
this leaks information regarding the number of \ofmaps. Finally, note that
every single byte of an \ifmap is read by a layer --
the total data volume (\textit{volume}) that is read is indicative of the number of \ifmaps. 
In fact, using these insights it
is possible to setup a system of equations and solve them to find all the layer parameters (see\cite{reverse}).} 

\blue{
Since the filters are read-only and never modified, the attacker
can easily distinguish between filter memory accesses and \ifmap/\ofmap accesses.
This leaks the size of the filter.
In addition to
tracking off-chip traffic, an adversary can monitor the execution \textit{time} of different
modules~\cite{cachetelepathy} to establish a correlation between the dimensions of the model and the
timing~\cite{cachetelepathy}. A recent attack \textit{HuffDuff}~\cite{huffduff} proposes a method
to craft specialized inputs such that the total data volume of non-zero values leaks the filter sizes. They use
a series of inputs that have one 1 and the rest 0s. 
}

\noindent \textit{\blue{Countermeasures -}} 
\blue{Recent works~\cite{neurobfuscator,mitigating,dnncloak} have proposed a set of
obfuscation strategies to conceal these leaked signatures.
They propose to shuffle accesses, widen the convolution layer by padding zeros,
increase the dimensions of the filters, augment the input with dummy data/zeros, 
buffer a part of the \ofmap 
or fuse/split layers. All of these methods increase the
model search space but fail to hide important signals such as the access count or
read-after-write dependence information (also pointed out in HuffDuff~\cite{huffduff}). }



%% file: Motivation.tex

\section{Motivation} 
\label{sec:motivation}


\subsubsection{$1^{st}$ Generation Statistical Attack (SS)}
\label{sec:SS}
Attackers can always collect multiple measurements across various runs, inputs and devices, and perform sophisticated analyses. As long as doing this is practically feasible, it will continue to be a method of choice because these attacks are easy to mount. 

\begin{figure}[!htb]
    \centering
    \includegraphics[width=0.84\linewidth]{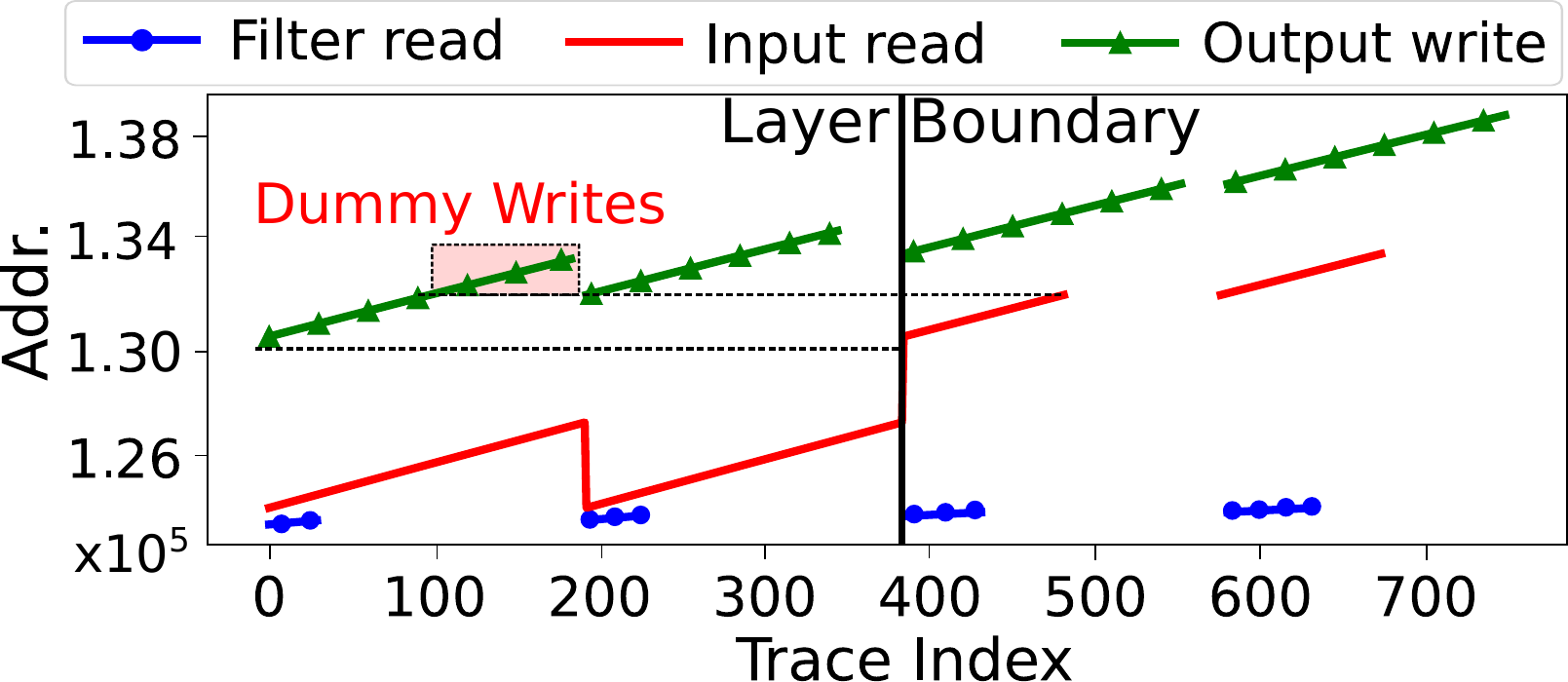}
    \caption{SS attack: Adding dummy writes (784) to obfuscate the true writes (1568) in the scaled version of the first layer of {\em Vgg16} ($32 \times 32 \times 3$).
    Adversaries eliminate the dummy writes as they are not read in the subsequent layer~(Liu et al.\cite{mitigating})}
    \label{fig:dummyWrites}
\end{figure}

For instance, when adding dummy writes (redundancy based noise~\cite{mitigating}) in a system, it is necessary
to read the dummy data, and then write something back. Otherwise, it is obvious that the original data is
ineffectual. \blue{Figure~\ref{fig:dummyWrites} shows an example of a CM \cite{mitigating} that had this vulnerability
and a method to detect and eliminate such writes}. 

On similar lines, \textit{shuffling} the addresses~\cite{mitigating} will also prove to be ineffective, as they also preserve the average read-write (RW) distance~\cite{dnncloak}. 

\begin{stbox}{Can we stop statistical attacks?}
 The adversary will estimate the pattern in the leaked data or parameters of the noise distribution by analyzing multiple samples collected over space and time.
\end{stbox}

\subsubsection{$2^{nd}$ Generation Kerckhoff Attack (KK)}
\label{sec:KK}
It is possible that even after collecting multiple samples, an adversary cannot filter out the noise or estimate
parameters of the distribution such as the value of the non-zero mean or the minimum. The $2^{nd}$ generation \textit{Kerckhoff}
attacks are built over the $1^{st}$ generation attacks to leak
such \blue{hardwired} parameters.
It is possible for a \blue{malicious insider} 
to leak them in accordance with the Kerckhoff's principle. 

\begin{figure}[!htb]
    \centering
    \includegraphics[width=0.8\linewidth]{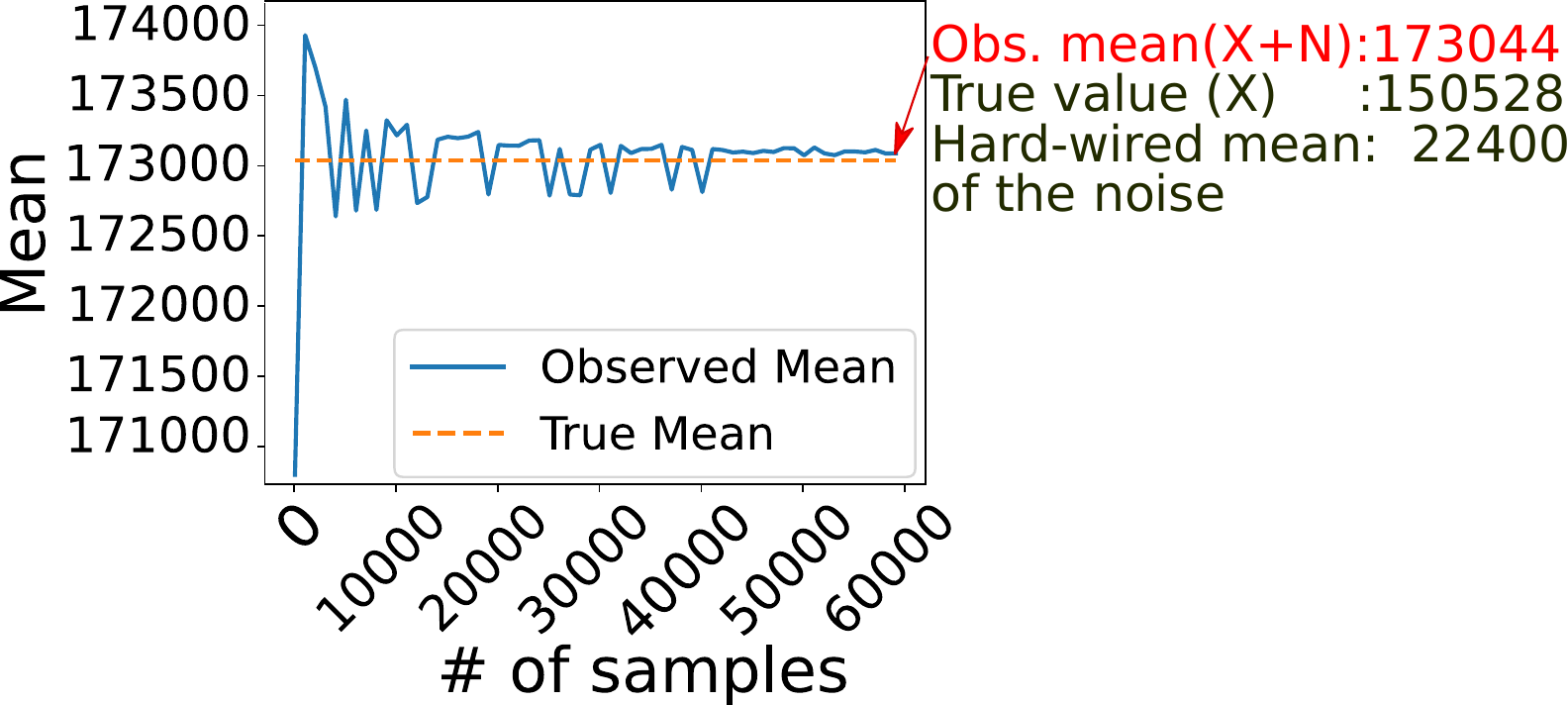}
    \caption{KK attack: A noise with a hardwired mean value of 22400 is added to obfuscate the actual volume of the first layer of {\em Vgg16} ($224\times 224\times 3$)~\cite{neurobfuscator,obfunas}. The adversary will eliminate the noise using an SS attack followed by the \textit{Kerckhoff}'s (KK) attack}
    \label{fig:diffMean}
\end{figure}

Some popular approaches~\cite{neurobfuscator,obfunas} obfuscate the layer size by adding dummy accesses (\textit{redundancy in space}), while DNNCloak~\cite{dnncloak} removes the accesses by buffering the data on-chip (\textit{removal}). Let us assume that the size of the layer is $X$ and the noise is $N$. To counter this, an adversary first conducts a statistical attack by collecting a multitude of data samples. They will converge to the mean as per the law of large numbers
(shown in Figure~\ref{fig:diffMean}). In this case, the noise added is 22400, which is 
a hardwired constant and can be leaked. Even if it is tied to the plaintext or the hardware (via 
PUFs~\cite{puf}), taking multiple samples will ultimately produce the mean and the minimum
(the most useful parameters). 
  
\begin{stbox}{Do not violate the {\em Kerckhoff} principle}
    Any hardwired secret parameter or constant may be leaked by a \blue{\textit{malicious insider}}. 
\end{stbox}

\subsubsection{$3^{rd}$ Generation Residual Side-channel Attack (SI)}
\label{sec:SI}
To circumvent the KK attack, we can make many of the hardwired constants a part of the
{\em key}. 
\blue{The advantage of this method is that a malicious insider cannot leak them, and the key
can be changed as frequently as we want
(see ~\cite{securelease}).}
An adversary always has some idea of the nature of the neural network that is being used based on what she knows from prior work and the nature of the problem that is being solved -- it would be
unwise to assume otherwise. This information can be weaponized. 

\begin{figure}[!htb]
    \centering
    \includegraphics[width=0.8\linewidth]{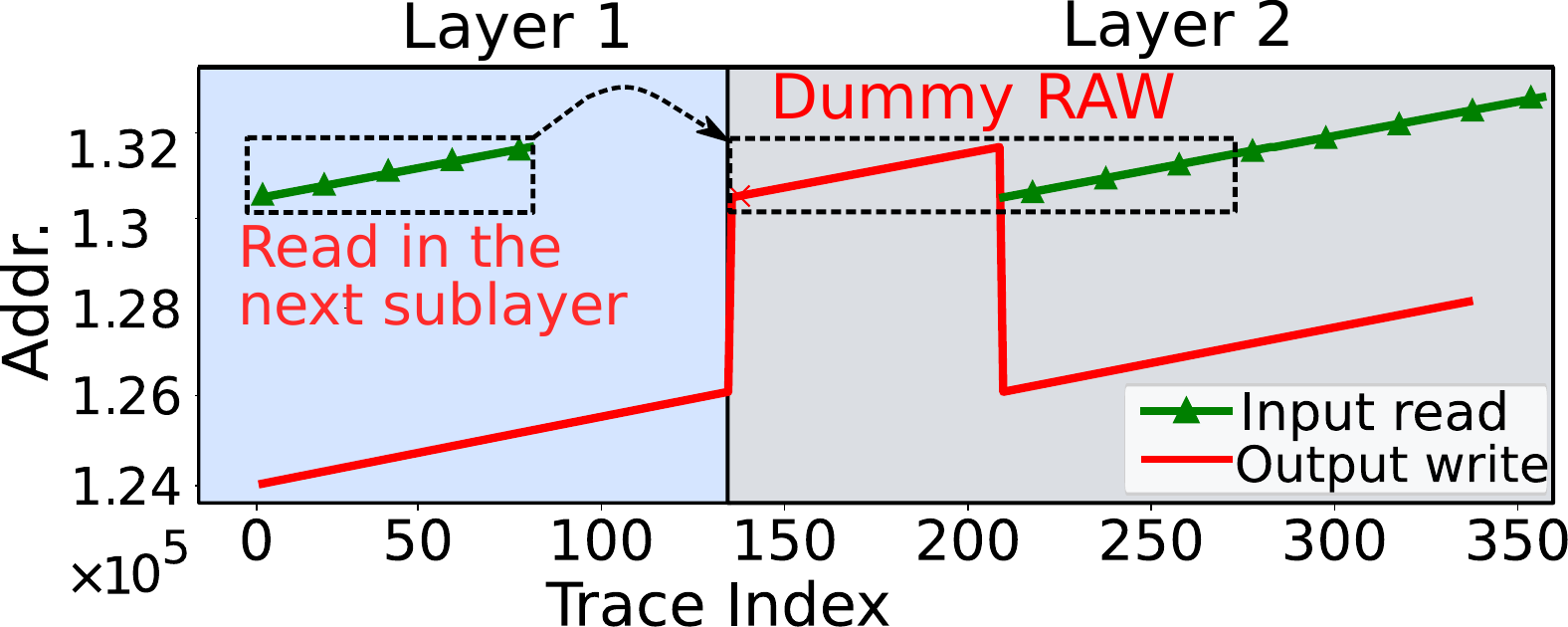}
    \caption{SI attack: Layer divider technique~\cite{dnncloak} to obfuscate RAW dependencies. The adversary filters out the fake RAW \blue{dependences} by checking the unchanged values.}
    \label{fig:LD}
\end{figure}

Consider the example shown in Figure~\ref{fig:LD}. It shows an execution snippet from DNNCloak
(Che et al.~\cite{dnncloak}). Here, a layer is split into sublayers and dummy accesses 
are added to the sublayers.  The idea is to
read the output of the previous sublayer into the 
current sublayer and write the output of both the
sublayers simultaneously (current and previous). This is done with the hope that dummy writes
cannot be filtered out because they are read and used in the next sublayer -- artifical RAW
dependencies are being created \blue{as} everything that is being written is being read. 
The adversary can {\em check} the contents of the addresses 
that are getting updated. As shown in the figure,
it is easy to spot that the values of some addresses \blue{that do not get updated}. This will
still be the case \blue{even if the data as well as the addresses} are encrypted with the same key. Based on an analysis
of access patterns in state-of-the-art DNNs, any adversary will know that the occurrence of this
phenomenon is highly unlikely and thus they are dummy writes in all likelihood. Even if we shuffle
the addresses, this pattern can still be discerned for individual blocks of addresses.

To reduce the search space, we can use another trick. We analyzed all the contemporary neural networks.
By and large, each layer is of the form $M^2\times N$, where $M$ is a large integer and $N$ is comparatively
smaller for the first few layers. Such numbers are known as NSQF (non square-free) numbers. If we only
consider NSQF numbers~\cite{squareful} 
with a set of constraints derived from layers' dimensions in contemporary DNNs,
the search space can reduce considerably. 

\begin{stbox}{There is always a residual side-channel}
An adversary can collect side-information from contemporary DNNs and use it to estimate the parameters of the target model and reduce the search space.
\end{stbox}


%% file: theory.tex

\section{Theoretical Analysis}
\label{sec:theory}

Consider a random variable $Y$, which corresponds to \blue{the number of tiles that the
attacker {\em observes} for a given layer}. This is related to the actual number
of tiles $X$, the compression ratio $C$ and the amount of noise $N$
as $Y = CX + N$.

Given that $C$ is a compression ratio, the following constraint holds: $0 < C \le 1$. 
The aim is to arrive at $X$ from $Y$.

\subsection{\blue{The Additive Noise: $N$}}
Any adversary would like to reduce the degrees of freedom and keep $C$ constant such that
the effect of \blue{the noise $N$ (dummy data)} can be eliminated. This can be ensured by making minor changes to the 
input such that the neural network's layers effectively filter them out. For example, we can
add speckled noise that does not correspond to any identifiable feature~\cite{noise}.
Multiple measurements will yield $CX + N_{min}$. 
Here, $N_{min}$ is the minimum value that $N$ can take. If
$N_{min}$ is known (as per the \textit{Kerckhoff}'s principle),
then $CX$ is exposed. 

\blue{If an adversary can estimate $C$ (based on prior studies), the system's security is compromised.}
We can substitute $N_{min}$ with
any other statistical metric like the mean or variance -- it does not 
make a difference. As we discovered in our
experimental analyses, tying $N$ to the model input
or PUFs does not significantly help because it is possible to collect a distribution of $N$
anyway by running experiments on multiple devices or using different model inputs\blue{\cite{canales2024polynomial,carlini2020cryptanalytic}}.

For the sake of explanation,
let us use $N_{min}$ as the metric of interest.
To make a significant difference, it should not be derived from the
model input, it cannot be zero and it definitely cannot be {\em known}. 
If it arises from a distribution,
then we shall continue to have the same problem. The only ``Kerckhoff safe''
approach is to make the parameters that govern $N$'s distribution like
$N_{min}$ a part of the {\em key} (encrypted model in our case).

Even if $N_{min}$ is a part of the key, we are still allowing the attacker to guess it with a finite number of choices. Hence, it is a good idea to create a distribution on top of it just to increase the computational difficulty.
Let us thus modify our equation as $Y = CX + \alpha + N'$.
In this case, $\alpha$ is a constant (should be a part of the key, proxy for $N_{min}$) and $N'$ is a distribution whose support is from $0$ till $R$, where $R$  along with the
parameters of the distribution should also be a part of the key.  Let us again give
the adversary the benefit of doubt, and assume that she can isolate measurements where $N'=0$.
In this case, $Y = CX + \alpha$. 

\subsection{\blue{The Multiplicative Noise: $C$}}
Let us assess the difficulty of guessing $C$.
Assume that the adversary knows some distribution of $C$ ($f(C)$) for a given layer 
(learnt from other DNNs). Let us give
the adversary the benefit of doubt and assume that she knows when a layer begins and ends.

Having some knowledge of $f(C)$ gives a good starting point to search for $X$. The problem
is finding the actual compression ratio $\beta$. In this case,
$Y = \beta X + \alpha$. Unlike the noise $N$, 
statistical and \textit{Kerckhoff} attacks are hard to mount on the
compression ratio mainly because of two reasons. The first is that it cannot be 
a hardwired constant and thus hardware designers will not know what the compression ratio for
a given layer will be. 


\blue{In the case of the additive noise, a standard approach is to conduct measurements using a diversity of inputs~\cite{carlini2020cryptanalytic,canales2024polynomial} and
find the mean or minimum value of $X+N$. With multiplicative noise of the form $CX$, an analogous metric is the
geometric mean. It is not very helpful mainly because either $C$ does not change with a perturbed input
or it does not change in a way
that is known to the attacker. 
The latter is because of \textit{aleatoric} uncertainty, which comes from randomness in the data,
and \textit{epistemic} uncertainty, which arises from limited knowledge about the model. Because of these uncertainties,
the only extra information that the attacker can have is a prior distribution $f(C)$ (learnt from
other models that may be similar). This is side information and its quality depends on how well the model
under attack is similar to known models in the literature. Note that $C$ can always be \underline{modulated} by the
model creator by adding random dummy data in a layer. The amount of dummy data to be added can be a part of the key.}

\subsection{Predicted Distribution of $X$}
The actual distribution of $X$ is not known to the adversary. It however knows the following
formula without knowing the values of $\alpha$ and $\beta$.
\begin{equation}
{X} = (Y-\alpha) \times \frac{1}{\beta}
\end{equation}
$Y$ is the value that the adversary observes, \blue{which in our design will be the DRAM traffic volume
for compressed and then encrypted data}. 
We are basically looking at a product of two random variables here. 
Let the random variable $U$ denote $Y- \alpha$ and $V$ denote $\frac{1}{\beta}$.
We have:
\begin{equation}
X = UV
\end{equation}
If we take the Mellin transform~\cite{mellintr} of both sides, we get. 
\begin{equation}
\mathcal{M}(X) = \mathcal{M}(UV) = \mathcal{M}(U) \times \mathcal{M}(V) 
\end{equation}
A Mellin transform is defined as: 
\begin{equation}
g(s)= \mathcal{M}(f(x)) = \int_0^\infty x^{s-1}f(x)dx
\end{equation}
The inverse transform is: 
\begin{equation}
f(x) = \mathcal{M}^{-1}(g(s)) = \frac{1}{2\pi i} \int_{c-i \infty}^{c + i \infty} x^{-s}g(s)ds
\end{equation}

Given a $Y$, an adversary can thus get a distribution of $X$ using an estimate of the
distributions of
$\alpha$ and $\beta$ ($f(C)$), and then by using {\em Mellin} transforms.
Let us represent the adversary's predicted distribution by the pdf $h(X)$. 

\subsubsection{Difficulty of Verification}
Given a distribution of $X$, the most optimal way of making guesses is to first choose the
most probable value (supremum), then the
second most probable value, and so on (see Massey~\cite{massey} for the proof).
Let us thus define the $rank$ function
that takes the \blue{adversary's predicted} distribution $h(X)$ and the real value of $X$ ($X_r$) as input. It sorts
the values of $X$ in descending order of their probabilities and returns the rank of $X_r$.
For example, if $X_r$ is the mode of the distribution, its rank is 1. Hence, we have a way
of \blue{\textbf{quantifying the search space}}. If there are $N_i$ choices in layer $i$, then the
total number of choices is $\Pi_i N_i$. Note that layers' dimensions/parameters in this approach
cannot be independently verified. 

\blue{This creates an all-or-nothing scenario, requiring the verification of a predicted combination of dimensions for the entire neural network in one comprehensive step. This is done by training and verifying the accuracy of the predicted network.}

\subsubsection{Imperfect Knowledge of $f(X)$}
The \textit{Mellin}
transform's computation involves approximating the integral using methods like the Riemann sum\cite{mellin}. 
However, its quadratic
complexity  imposes limitations on its practical utility for extensive computations~\cite{mellin}. Sena et
al.~\cite{mellin} proposes an alternative that uses
the FFT to compute the \textit{Mellin} transform. This is achieved by 
performing spline interpolation, exponential resampling, exponential multiplication followed by FFT computation. 

The adversary uses this approach to \blue{estimate the}  distribution $h(X)$. 
Next, she performs an interpolation based on the
\textit{piecewise cubic hermite polynomial (PCHIP)}\cite{pchip}, to modify $h(X)$
such that the probability of all non-NSQF numbers is zero. Every NSQF number
accumulates the probabilities of all the non-NSQF numbers in its neighborhood
(half the distance on either side to the nearest NSQF number). This is our 
{\em smart search space} (probability distribution: $h'(X)$). 

\noindent \textbf{Rank Estimation} We ran this experiment for estimating the rank of $X$ (layer volume)
for the first five layers of \textit{Vgg16} 
(see Figure~\ref{fig:rank}), where the true value of \ifmap ($X_r$) for each layer is 150528, 802816,  401428, 802816, and 200704, respectively. 


In order to estimate the side-information available to the adversary regarding the compression ratio, we studied the compression ratios for nine \textit{cognate} open-source DNN models -- {\em AlexNet}~\cite{alexnet},
{\em Vgg16}~\cite{vggnet}, {\em Vgg19}~\cite{vggnet}, {\em SqueezeNet}~\cite{squeezenet}, {\em ShuffleNet}~\cite{shufflenet}, {\em ResNet}~\cite{resnet},
{\em MobileNet}~\cite{mobilenet}, {\em Googlenet}~\cite{googlenet} and {\em Inception} Net~\cite{inception}. 
Based on the observed range of the compression ratios (1.5$\times$-- 40$\times$) (also observed in~\cite{deepCompression}) across different DNN models (also pruned and quantized), we generate different distributions for $U$ and $V$. 
We vary $\alpha$ between [100, 224 $\times$ 224 $\times$ 128 (layer size)]. Next, we compute $h'(X)$ using our
Mellin transform based approach.
We observe that the uniform distribution has the highest
rank of $X_r$ followed by the geometric distribution. 

\begin{figure}[!htb]
    \centering
    \includegraphics[width=0.9\linewidth]{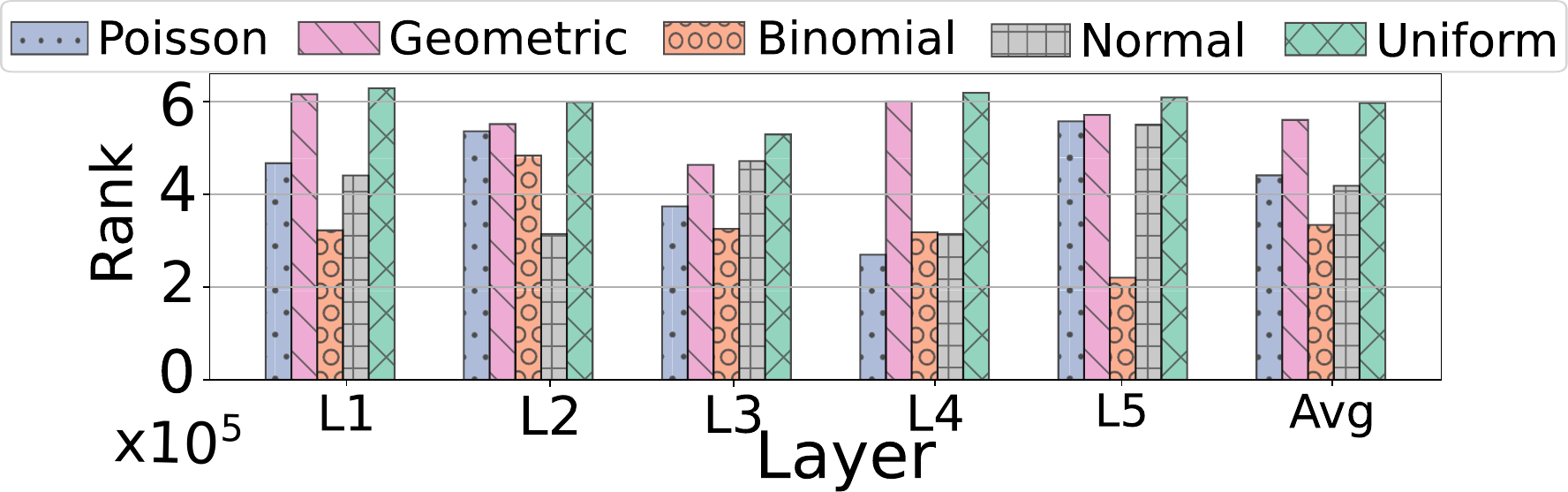}
    \caption{Rank of $X_r$ (layer volume) for different layers of \textit{Vgg16}. From our point of view the
    uniform distribution is the best. }
    \label{fig:rank}
\end{figure}

%% file: design.tex
\section{Design of NeuroPlug}
\label{sec:hw}
\vspace{-3mm}
\subsection{Overview}
\begin{figure}[!htb]
    \centering
    \includegraphics[width=0.9\linewidth]{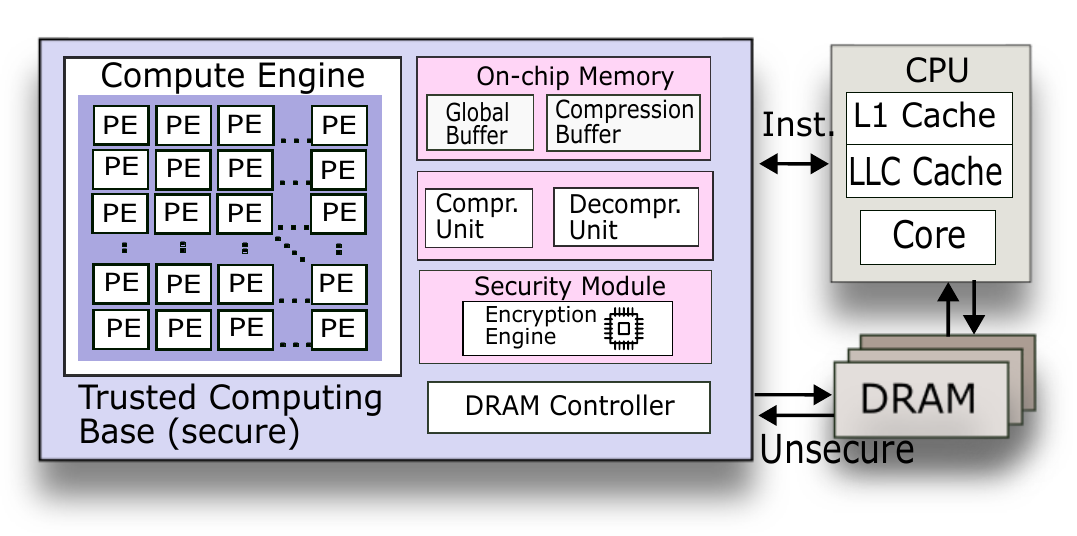}
    \caption{The high-level design of \fname}
    \label{fig:scheme}
\end{figure}

The high-level
design of \fname is shown in Figure~\ref{fig:scheme}. The host CPU securely delivers instructions to the
NPU via a PCIe link to execute a layer of the DNN. Each layer is divided into a set of tiles 
(basic unit of computation).  Subsequently, the NPU starts tile-wise computation of convolutions and other
linear/nonlinear operations.

We map the complex multi-dimensional computation space of a DNN's layer to a 1-D space using a SFC (space-filling curve). 
The SFC conceptually moves through all the tiles in a layer -- first along the channels, then
along the rows (see Listing~\ref{lst:conv}).  
In the SFC, we compress a sequence of contiguous tiles and fit them in a {\em bin}, 
where all the bins have the same size. 
Tiles can be split across
bins and we can also leave space empty within a bin. A {\em Bin Table} is stored at the 
beginning of a bin that 
stores the starting address of each compressed tile in a bin. 

\begin{figure*}[!htb]
    \centering
    \includegraphics[width=\linewidth]{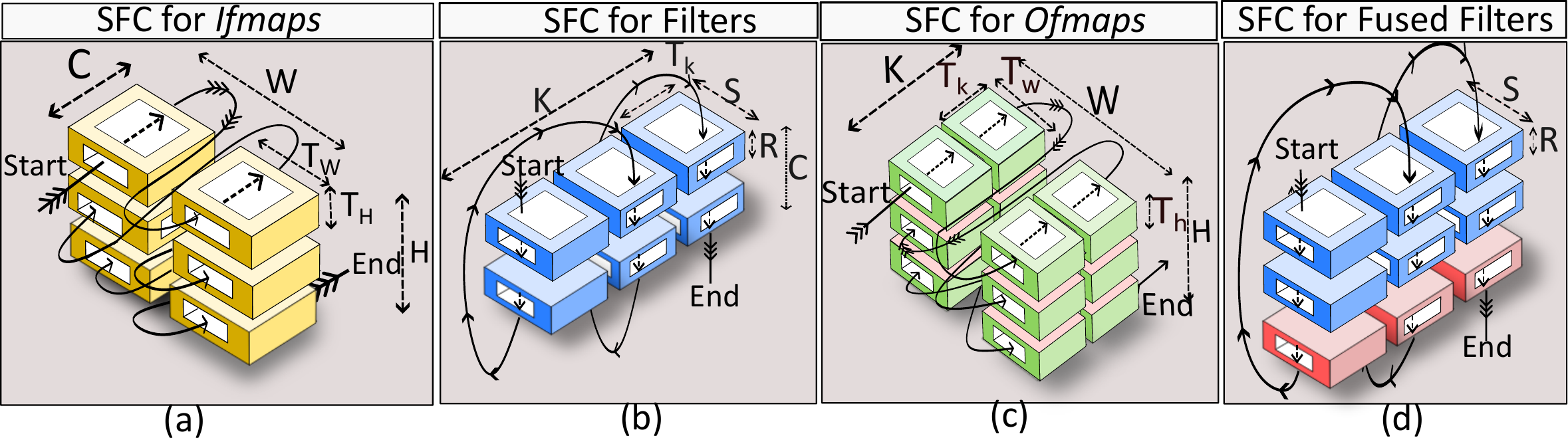}
    \caption{SFCs ({\textbf{a}}) Reading \ifmaps; (\textbf{b}) Reading the filters in a layer; (\textbf{c}) Writing \ofmaps; 
     (\textbf{d}) Reading fused filters across two layers}
    \label{fig:SFCs}
\end{figure*}

The {\em bin} is the atomic unit of {\em storage}
in our system, whereas a {\em tile} is the atomic unit of {\em computation}. 
A bin is read and written in one go. Furthermore, we ensure that it takes exactly the same amount of time to
process each bin such that timing side-channels are plugged. We place a bound $\kappa$ on the maximum number of tiles
that can be stored within a bin -- the processing time of a bin
corresponds to processing all $\kappa$ tiles (included in the
calculation of performance overheads). 

The compute engine is a traditional systolic PE array; each PE has
local storage in the form of register files and local buffers and a tile decompression unit. 
As the \textit{first layer is the most exposed}, we enhance its security by initially packing
dummy data within the tiles before performing compression -- its location is stored in the Bin Table.

A dedicated compression engine~\cite{deepCompression} compresses the tiles, packs them in bins,
leaves some space empty (as per the
noise distribution) and writes them in a sequential order to the main memory (as an SFC for the next layer).
{\em This ensures that writes are also at the granularity of bins and follow the same semantics.} 
SFCs provide a convenient abstraction for representing computations within a DNN along with providing
performance benefits in the form of enhanced spatial locality.

\vspace{-2.4mm}
\subsection{Deep Look at SFC-based Computation}
Let us analyze the computation patterns with SFCs (refer to Table~\ref{tab:symbol} for the
terminology). In a simplistic avatar of our design, we assume that all the \ifmaps and
weights fit on chip (will be relaxed later). 
The \ifmap is stored as a simple 1-D 
SFC as shown in Figure~\ref{fig:SFCs}(a). Each cuboid represents a {\em deep tile} --
same tile across all the input channels. 
We start with the first deep tile (iterate across all the channels)
and then move to the next deep tile in the same row and so on. Once we finish processing a row,
we move to the next row (first column), so on and so forth. The weights are also stored as an 
SFC. A single 1-D SFC contains $CK$ kernels -- we first read all the weights to generate \ofmap 1,
then \ofmap 2, so on and so forth (refer to Figure~\ref{fig:SFCs}(b)). 
The \ofmaps have exactly the same format as \ifmaps -- this needs to be the case because they
are inputs to the next layer. Refer to Figure~\ref{fig:SFCs}(c) that shows the order in which we
write. Table~\ref{fig:principles} outlines the key design principles using SFCs. Let us now make our design
more realistic.

\begin{table}[!htb]
    \centering
    \begin{tabular}{|p{0.85\linewidth}|}
    \hline
     \rowcolor{gray!10}
     $\blacktriangleright$ Read/write  an \ifmap or \ofmap only once. \\
     \rowcolor{gray!10}
     $\blacktriangleright$ Read all the channels of an \ifmap tile together (a deep tile). \\
    \hline     
    \end{tabular}
    \caption{Design principles \label{fig:principles}}
\end{table}

\subsubsection{{\bf Case I}: Weights do not fit but input maps do}
Consider the case where the weights \blue{do not fit in the NPU memory buffers but the inputs do.}
In this case, we partition the weights on the basis of a partitioning of \ofmaps.
This is akin to partitioning the set of $K$ \ofmaps into a sequence of \blue{contiguous subsets} of
$k_1, k_2, \ldots k_n$ \ofmaps where $\sum_{i=1}^n k_i = K$. In practice, this is not visible
to the attacker because we read the filters for the first $k_1$ \ofmaps and compute them
fully, then do the same for the next $k_2$, up till $k_n$. The \ofmaps are stored one after the
other -- exactly in the same format as they would have been stored had all the weights fit
in memory. The timing of filter reads could be a side-channel, however, we obscure this
by ensuring that \blue{each bin takes a fixed amount of time to execute} and the inter-bin and inter-filter-bin
read time is the same. 

\subsubsection{{\bf Case II}: Input maps do not fit but weights do}
We split an \ifmap bin-wise such that each set of bins fits within the NPU (row-major order).
Then the deep tiles
in each set of bins are processed and the corresponding output tiles/bins are written to memory. This is
indistinguishable from the case in which all inputs and weights fit in on-chip memory.

\subsubsection{{\bf Case III}: Both inputs and weights do not fit}
We partition an \ifmap's SFC bin-wise such that each partition fits within the NPU. Then we proceed
on the lines of Case I (single pass through all the weights). We need to subsequently read the next set of
\ifmap bins and re-read all the filters once again. This is where a side-channel gets exposed. 
\blue{The frequency at which the same filter weights are read correlates directly with the total volume of input data. A straightforward solution is to unroll the weights. For instance, consider partitioning the weights into two sets: $\mathcal{W}_1$
and $\mathcal{W}_2$. }
We create an unrolled SFC of the form $\mathcal{W}_1 \mathcal{W}_2 \mathcal{W}_1 \mathcal{W}_2 \ldots$.
To save space, we can perform limited unrolling, where if we need to read the weights $\tau$ times, we set the
unrolling factor to $\eta$ (model-specific secret). In extreme cases, we can have a situation where a
deep tile does not fit in the NPU's on-chip memory; this case can be solved by writing to multiple small 
SFCs and reading them in an interleaved fashion in the next layer. {\bf Note that in all cases, the 
partitioning factor is a random variable that is generated in the same fashion as $N$}

\subsection{Dealing with Halo Pixels}
{\em Halo pixels~\cite{book} } \blue{are the} additional pixels that are required to compute the convolution
results in a tile (beyond its boundaries). In Figure~\ref{fig:halo}, the halo pixels will be towards the north and west. 
\begin{figure}[!htb]
    \centering
    \includegraphics[width=0.5\linewidth]{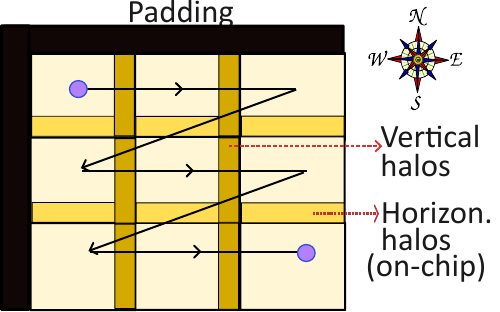}
    \caption{Halo pixels in a layer}
    \label{fig:halo}
\end{figure}
Let the processing begin at the Northwest corner. 
We then proceed along the same row (towards East). Each tile in the northernmost row get its halo pixels
from its western neighbor that has already been read. These pixels can be stored in a small on-chip buffer
for immediate use. The number of halo pixels is quite small 
as compared to the sizes of the
tiles. We proceed till we reach the last element of the row (eastern most). 
We then start at the westernmost column of the next row, and proceed eastward. In this case, the halo pixels
need to come from the western neighbor and northern neighbor. The western neighbor (tile) was just read,
however the northern neighbor (tile) was a part of the previous row. The first option is to store all the halo
pixels in an on-chip buffer. In this case, when we are processing a row,
we need to store the halo pixels for the southern neighbors in an on-chip buffer. However, if there are space
overflows, then we can write them as an SFC, and read them back in the same order when we are processing the
next row.

\subsection{Adding Custom Noise}
The uncertainty in the number of tiles in each bin is integral to 
maintaining the overall 
security of the design as it obfuscates the \textit{total data volume} 
and the \textit{RW distance}. \blue{In $CX+N$, $C$ naturally arises from
compression and it can further be augmented with adding random dummy data~\cite{nissim2007smooth,geng2015optimal}.}
The additive noise $N= \alpha + N'$ can be added by keeping $N$ bytes empty within a bin. 
\blue{For computing $N'$, we sample it from a heteroskedastic noise distribution, which
makes the adversary's job even harder. }
\blue{This distribution is characterized by its non-deterministic variance.}
This makes any kind of regression-based analysis hard. 
We performed the White~\cite{whiteTest} and Breusch-Pagan~\cite{BPTest} tests using different combinations of distributions
and estimated that a strong heteroskedastic noise distribution can be generated using a combination of the uniform and
Gaussian distributions -- both have very efficient hardware
implementations~\cite{uniform1,uniform2,gaussian1,gaussian2}. We use the uniform distribution(choice supported
by~Figure~\ref{fig:rank}) to set the variance of the Gaussian distribution. 

\subsection{Layer Splitting and Fusion}
In our architecture, \blue{performing} layer splitting~\cite{neurobfuscator} and fusion are easy. In layer splitting,
we partition an \ifmap and assign each part to a sub-layer.
In this case, 
the input SFC will be read once but the weight SFC may be read multiple times. We need to adopt
the solution proposed in Case III. For layer fusion, we can split the \ifmap if there is an on-chip
space issue. Then for each set of deep tiles, we need to read the filters
for two consecutive layers one after the other (Figure~\ref{fig:SFCs}(d) shows how
the filters are fused). For handling issues concerning limited
on-chip storage, we can adopt the solutions for Cases II and III.

\subsection{Pooling, ReLU and Skip Connections}
The pooling and ReLU operations
can be performed 
along with the convolution layer. 
Note that the tile size will change 
if we are pooling. If we are replacing a pixel with the maximum value in a window
of 4 pixels, then the tile size will reduce by a factor of 4.
In case, the
tile size reduces to a very small value (let's
say $1 \times 1$), then we assume enough buffer space to read a large part of the input,
recreate tiles, compress and place them into bins.
Our design can easily handle skip connections as well --
it simply requires re-reading the SFC of a previous layer. This will convey to the
adversary that a skip connection exists (limitation in our design), however because
of layer-boundary obfuscation, finding the source of the skip connection will be hard.

%% file: evaluation.tex
\section{Performance Evaluation}
\label{sec:eval}

\subsection{Setup and Benchmarks}
Our experimental setup  integrates the state-of-the-art DNN modeling tool (\textit{Timeloop}~\cite{timeloop})
and a cycle-accurate DRAM simulator (\textit{Ramulator}). Both the simulators have been validated against real-world HW and
are widely used in the literature. \textit{Timeloop}~\cite{timeloop} takes as input the details of the workload, the target
NPU's configuration (see Table \ref{tab:acc_config}), and the data reuse strategy as the input and generates
performance statistics and traces. 

We developed an in-house trace generator tool that extracts the memory access patterns of tiles from \textit{Timeloop's}
T-traces~\cite{Ttrace} and then compresses and randomizes the tiles in accordance with our
compression and binning strategies. Then, the tool generates a new sequence of memory accesses corresponding to bins.
\textit{Ramulator} takes traces from our trace generator tool as its input and generates the performance statistics. 


\begin{table}[!htb]
    \centering
    \footnotesize
    \begin{tabular}{|l|l||l|l|}
    \hline
    \rowcolor{black!10}
     \textbf{Parameter} & \textbf{Value} & \textbf{Parameter} & \textbf{Value} \\
    \hline
     PE size & 16 $\times$ 12 & Comp. buffer & 182 kB\\
     PE register & 440 B & Global buffer & 182 kB \\
      DRAM & DDR3 &  Frequency & 1.6 GHz \\
     \hline
    \hline     
    \end{tabular}
    \caption{Configuration of the NPU}
    \label{tab:acc_config}
\end{table}

\noindent \textbf{Workloads} We consider a set of dense as well sparse 
NN workloads (same as ~\cite{huffduff,neurobfuscator,dnncloak}). 
We generated pruned NNs using the layer-adaptive magnitude-based pruning technique~\cite{lamp} 
with a fairly good prediction accuracy as shown
in Table~\ref{tab:pruneAcc}. Note that \fname will work for all types of convolution-based DNNs such as GANs and Transformers)
\begin{wraptable}[10]{r}{0.42\linewidth}
\footnotesize
    \begin{tabular}{|l|l|}
    \hline
    \rowcolor{black!10}
    \textbf{Model} & \textbf{Acc.}\\
    \hline
    \rowcolor{gray!20}
     \multicolumn{2}{|c|}{\textbf{Pretrained}}\\
     \hline
      {\em ResNet18}~\cite{resnet}   & 89.1\% \\
      {\em Vgg16}~\cite{vggnet} & 91.32\% \\
      {\em AlexNet}~\cite{alexnet} & 81\% \\
      \hline
     \rowcolor{gray!20}
     \multicolumn{2}{|c|}{\textbf{Pruned}}\\
     \hline
     {\em ResNet18}~\cite{resnet}  & 92.1\% \\
     {\em Vgg16}~\cite{vggnet} & 90\% \\
     {\em AlexNet}~\cite{alexnet} & 80\% \\
      \hline
    \end{tabular}
    \caption{Workloads}
    \label{tab:pruneAcc}
\end{wraptable}
since other basic layers such as the fully-connected layer can be easily expressed as a convolution~\cite{securator}.
We are not showing the results for Transformers and LLMs due to a paucity of space.
Their sizes are huge and mounting a physical attack to estimate the model parameters is very hard (massive search space).
The main target of such attacks are smaller DNNs.

We use an NVIDIA Tesla
T4 GPU with CUDA 12.1 to train and prune the NN models using the CIFAR-10 dataset. 
We achieved
an average accuracy of 90\% for {\em Vgg16}, 80\% for {\em Alexnet} and 92.1\% for {\em ResNet18} -- same as the original authors. During our analysis, we observed that many other popular CNN models demonstrated a similar
behavior (not discussed due to paucity of space). 

\noindent \textbf{Configurations}  We select an unsecure system as the {\em baseline} configuration. The
bin size was set to 60 kB. The TCB can store three bins at a time. 
We observed that the tile size varied with the layers and was roughly between 10 kB - 95 kB. We assume that an adversary analyzes the compression ratio ($\beta$) using other sister DNNs and crafts bespoke inputs; she observes that the values are normally distributed in the range $1.5-40\times$~\cite{deepCompression} (also seen in our experiments).
The {\em Compr} design represents a two-level compression (pruning~\cite{lamp} followed by {\em Huffman} compression) similar to the Deep Compression Algorithm~\cite{deepCompression}. The other configurations are -- DNNCloak {\em
(DNNCloak)}\cite{dnncloak}, Layer deepening ({\em NeurObf\_1})~\cite{neurobfuscator}, Layer widening ({\em
NeurObf\_2})~\cite{neurobfuscator}, and Kernel widening ({\em NeurObf\_3})~\cite{neurobfuscator}. In \fname, for the given set of models, the upper limit of the heteroskedastic noise distribution $\alpha$ was set to $8000$ (in bytes).

\subsection{Performance Analysis}
\begin{figure}[!htb]
    \centering
    \includegraphics[width=0.8\linewidth]{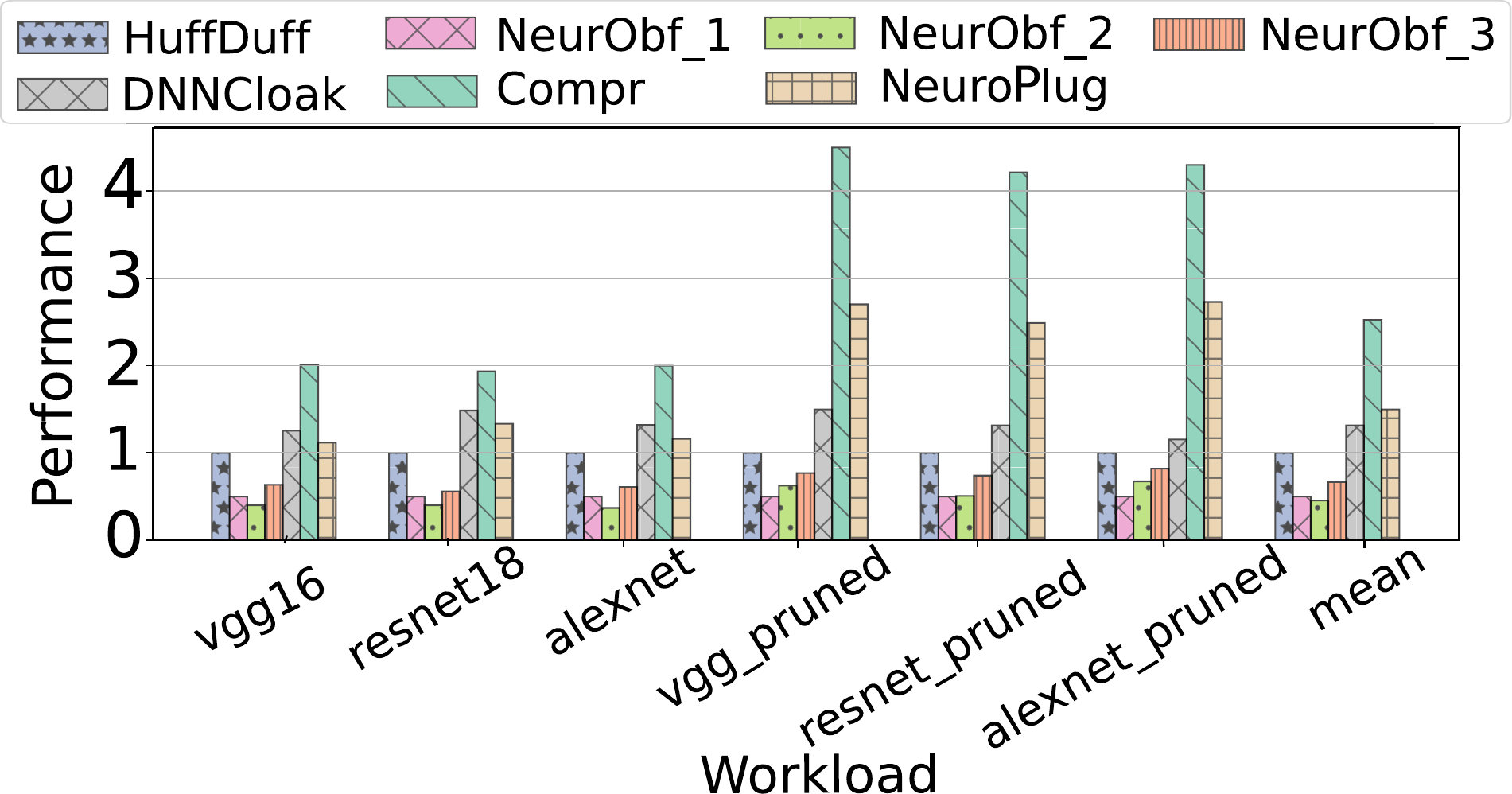}
    \caption{Normalized performance}
    \label{fig:perf}
\end{figure}

\begin{figure}[!htb]
    \centering
    \includegraphics[width=0.8\linewidth]{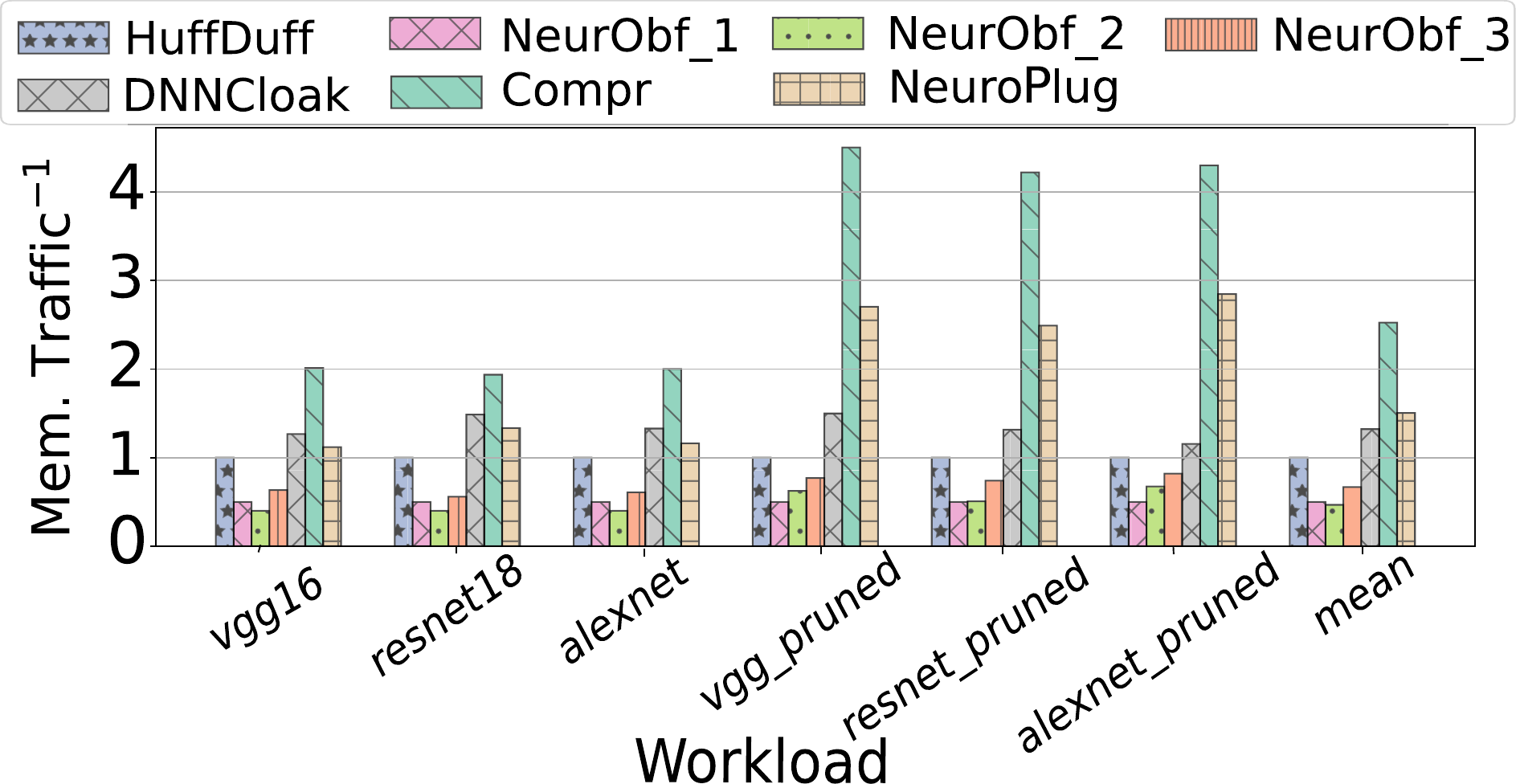}
    \caption{Normalized memory traffic}
    \label{fig:traff} 
\end{figure}

The performance results are
shown in Figure~\ref{fig:perf}. The {\em performance} is proportional to the reciprocal of the total \textit{\#simulation cycles}. The results are normalized relative to the
baseline setup. We find that {\em Compr} is $2.52 \times$ faster than the baseline. This is due to
the fact that compression leads to a reduction in memory traffic by a factor of $2.52 \times$ as shown in Figure~\ref{fig:traff}. There is a good correlation between performance and DRAM
traffic, which is aligned with the findings reported in prior studies ({\em TNPU~\cite{tnpu}}, {\em
Securator~\cite{securator}}, \textit{GuardNN}~\cite{guardnn}). 

We see that on an average, the rest of the proposals~\cite{neurobfuscator} are $1.8\times$ worse than  \fname in terms of performance. Note that
 {\em DNNCloak} performs \textit{only} weight
compression and incorporates an on-chip buffer to store the \ofmaps, thereby reducing the memory traffic. But, it also
includes dummy accesses and reads/writes the previous sublayer, which lead to an increase in the memory traffic, resulting in {\em DNNCloak} 
being $15.48 \%$
slower than \fname. On the contrary, \fname employs compression to reduce all the off-chip communication,
which allows us to afford more dummy accesses.

\subsection{Sensitivity Analysis}
\begin{figure}[!htb]
    \centering
    \includegraphics[width=0.8\linewidth]{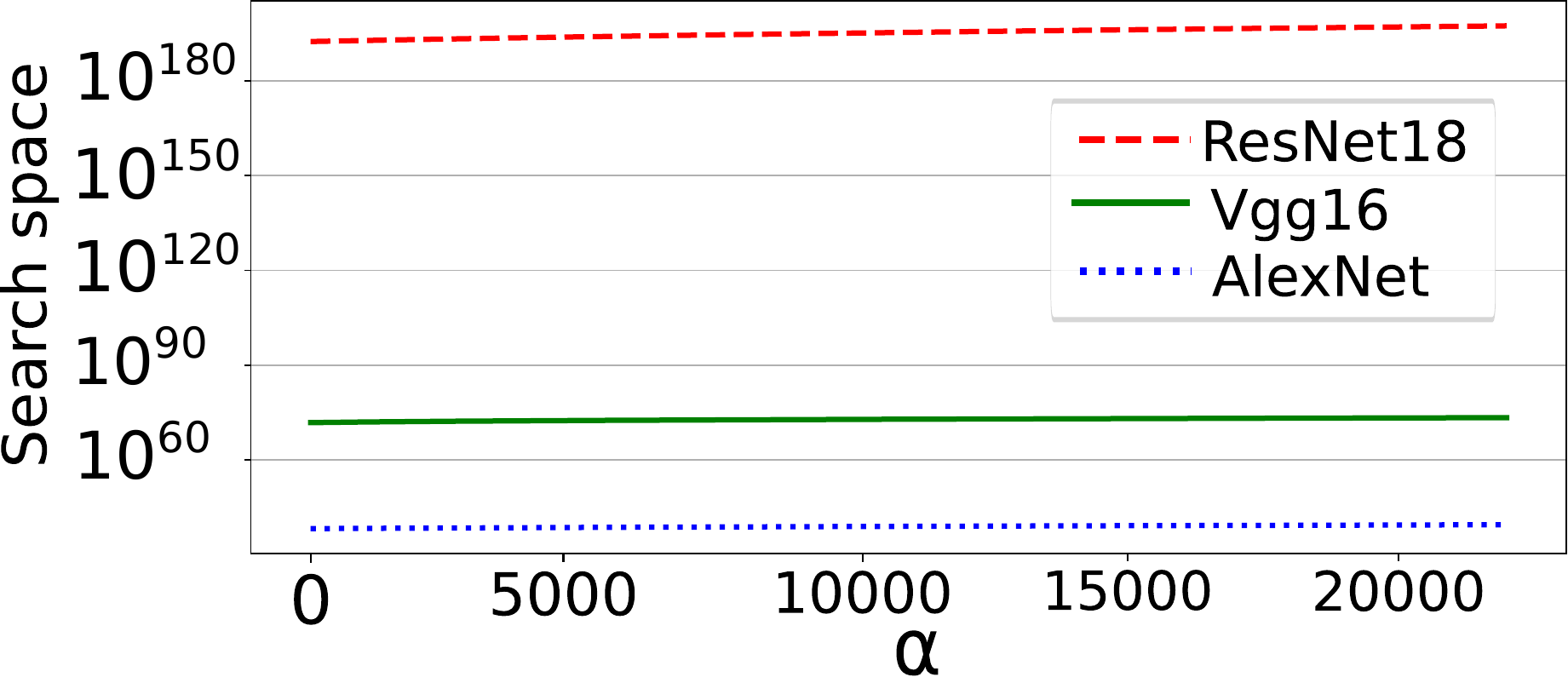}
    \caption{Noise sensitivity analysis}
    \label{fig:tradeOff}
\end{figure}

Figure ~\ref{fig:tradeOff} shows the variation in the smart search space with the 
noise level ($\alpha$) for \fname. 
The smart search space for {\em ResNet} is of the order of $10^{196}$, for
{\em Vgg16} is of the order of $10^{70}$ and for {\em AlexNet}, it is of the order of $10^{26}$ (small network,
deliberately chosen).
 
\begin{figure}[!htb]
    \centering
    \includegraphics[width=0.85\linewidth]{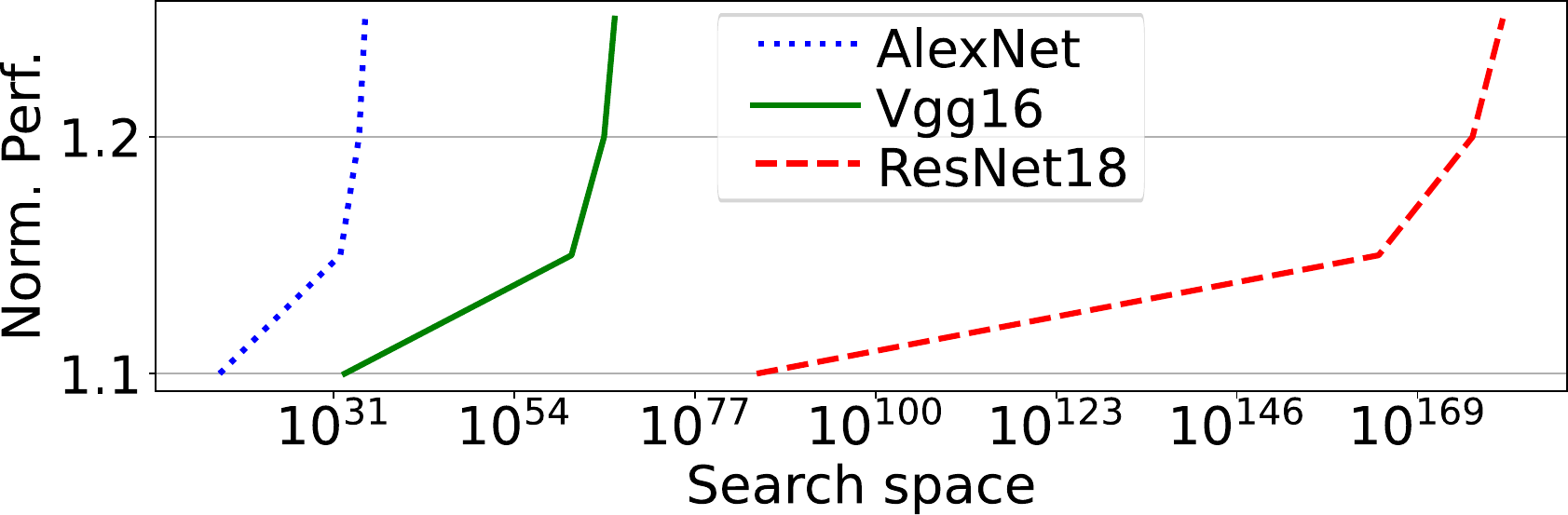}
    \caption{\blue{Performance overhead vs search space size ($\blacktriangleright$ overheads
    increase very gradually)}}
    \label{fig:SS-perf}
\end{figure}

\blue{Figure~\ref{fig:SS-perf} shows the trade-off between the search space size
and the performance overheads introduced by modulating $C$ (adding random values). 
}

\subsection{Hardware Overheads}
\blue{
\fname incorporates additional HW, which can be added to conventional NPUs.
The code for all the additional components was written in Verilog, and the designs were
fully synthesized, placed, and routed using the Cadence Genus tool for a 28 nm ASIC technology node at 400 MHz frequency. We present the results
in Table \ref{tab:asic}. Our design is generic and can work with any NPU (that works at the tile level) subject to making
minimal changes. We show the results for an NPU that is based on the Eyeriss-2 accelerator\cite{eyerissv2} (its area
is 5.5$mm^2$).
}

\begin{table}[!htb]
    \centering
    \footnotesize
    \begin{tabular}{|p{4cm}|c|c|}
    \hline
    \rowcolor{black!10}
    \textbf{Module} & \textbf{Area($\mu m^2 \times 10^{3}$ )} & \textbf{Power(mW)}  \\
    \hline
    Compr-encry. engines & 767 & 97.92 \\
    SFC addr. generator & 8.35  & 0.56  \\
    Binning logic & 1.26  & 1.47 \\
    Noise generator & 3.29  & 0.340 \\
   {Overall overhead (including buffers)} & \textbf{ 1581.40 } &\textbf{ 130.43 } \\
   \hline
   \rowcolor{gray!20}
    \textbf{Tool} & \multicolumn{2}{c|}{Cadence RTL compiler, 28 nm} \\
    \hline
    \end{tabular}
    \caption{\blue{ASIC area and power utilization for the additional hardware components used in \fname}}
    \label{tab:asic}
\end{table}

\blue{We also implemented the design on a Virtex-7 FPGA board (xc7vx330t-3fg1157). The number of LUTs, registers and DRAMs required to implement the additional logic are 84K, 77K and 8 (resp.) at a frequency of 360 MHz.}


%% file: securityEvaluation.tex
\section{Security Evaluation}
\label{sec:security}
We perform a thorough security analysis using multiple DNNs. We primarily
report the results for the {\em Vgg16} model, which is more vulnerable to attacks because of its lack of skip connections, good accuracy, and relatively small model size that
reduce the search space. The results for other NNs are similar or better (not presented due to space constraints).
\subsection{Search Space Sensitivity}
\begin{figure}[!htb]
  \centering
    \includegraphics[width=0.6\linewidth]{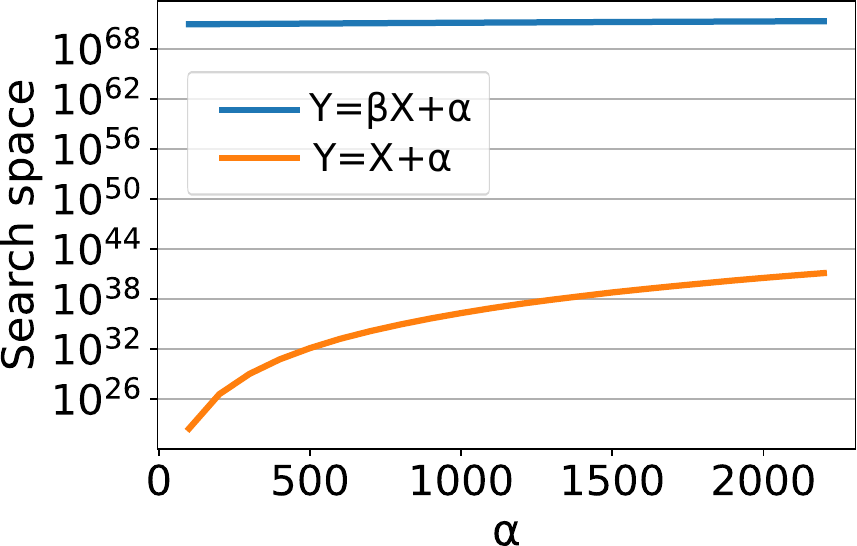}
  \caption{Variation in the smart search space size with and without compression (VGG16). {\bf Conclusion}:
  compression is needed.}
  \label{fig:SrS}
\end{figure}

We estimate the variation in the smart search space size as shown in Figure~\ref{fig:SrS} with different noise levels.
We vary the level of the noise $\alpha$ in the presence 
and absence of compression -- $Y = \beta X + \alpha$. Here $\beta$ is the ratio of the size of the 
compressed data and the size of the original data. 
We observe that with
the same level of noise($\alpha$), the search space of $Y = \beta X + \alpha$ is of the order $10^{70}$,
while for $Y = X + \alpha$, it is of the order $10^{26}$ (significantly smaller). 

\begin{figure}[!htb]
    \centering
    \includegraphics[width=0.8\linewidth]{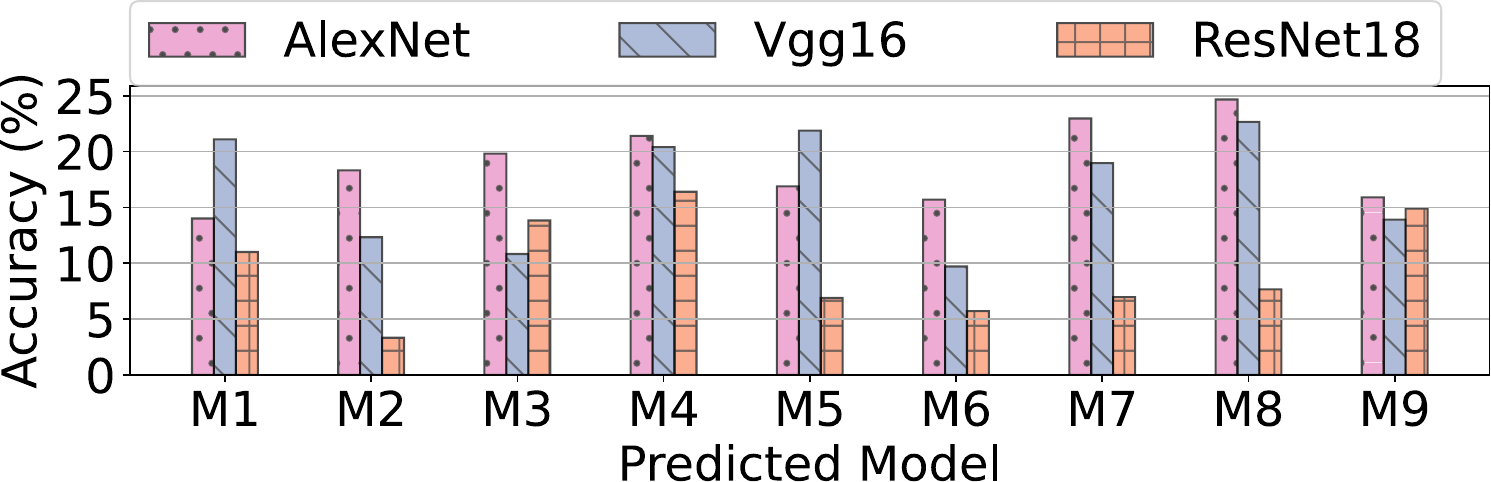}
    \caption{Accuracy of the models estimated by an attacker}
    \label{fig:retraining}
\end{figure}

We also mount an attack in the presence of \fname and report 
the accuracy of the {\em top} nine models in CIFAR-10 that can be estimated by 
the adversary in Figure~\ref{fig:retraining}. These accuracies are similar to the 
accuracy of a randomly generated model ($10-20 \%$), where no leakage hints are available.

Next, we perform two distinct case studies to examine the security of our scheme 
against two highly potent recent attacks:
\textit{HuffDuff}~\cite{huffduff} and \textit{Reverse Engg}~\cite{reverse}.
There is no known countermeasure for HuffDuff.
A total of 2048 unique memory traffic traces were collected using various inputs 
(as in the \textit{HuffDuff} model) in order to perform the security analysis for the 
\textit{Vgg16} model.
Note that the conclusions for other DNNs, such as \textit{ResNet18} and \textit{AlexNet}, are identical 
(not presented due to space constraints).

\subsection{Case Study 1 -- Defense from HuffDuff} 
We analyze the first five most vulnerable layers of the 
\textit{Vgg16} model
across three distinct configurations -- scheme without any CM, 
our proposed scheme (\fname) as the CM, and a hypothetical secure system {\em random}
that provides complete security guarantees, ensuring that any data observed by an adversary
is entirely random and unrelated to the architecture of the models. 
Other competing CMs are not considered since they cannot stop this attack.
We employ a combination of information-theory and statistics-based tests to 
assess the security of \fname. This approach is favored in the cryptography community 
as it avoids reliance on a single metric. For example, the classical correlation 
coefficient could be zero for random variables $x$ and $x^2$, however, 
other security metrics reveal their underlying connection. 

\noindent \textbf{Information Theoretic Tests -} We use FI (Fisher Information) and MI (Mutual Information), which are extensively used metrics to estimate the information
content of leaked signatures and their association with secret
data/keys~\cite{fisher1,fisher2,fisher3,fisher4,fisher5,mayer2006fisher,MIleakage,MIleakage2,MI3,MI4}.
{\bf These metrics can never be zero for a distribution with a finite support. Our aim is to bring them as close
to the random configuration as possible}.

\noindent $\blacktriangleright$ \textit{Fisher information (FI)}  
The FI is a robust and classical method for establishing a lower
bound on the error (variance) of the estimator -- the lower the better~\cite{fisher4}. 
We mount the {\em HuffDuff} attack and observe that the FI
is quite low for the first 5 layers (comparable to random). It reduces as we consider deeper layers
(due to epistemic uncertainty).

\begin{figure}[!htb]
  \centering
  \includegraphics[width=0.65\linewidth]{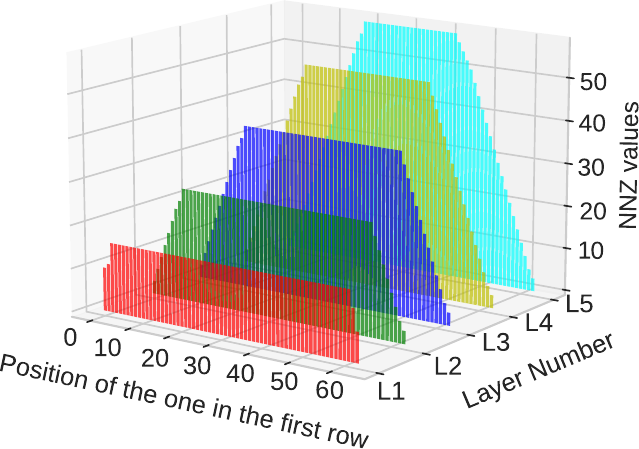}
  \caption{The boundary effect decreases with the layer depth. 
  64 different $64 \times 64$ inputs are created by varying the 
  position of the one in the first row of the input with all zeros. }
  \label{fig:huffduff}
\end{figure}

\begin{figure}[!htb]
  \centering
  \includegraphics[width=1\linewidth]{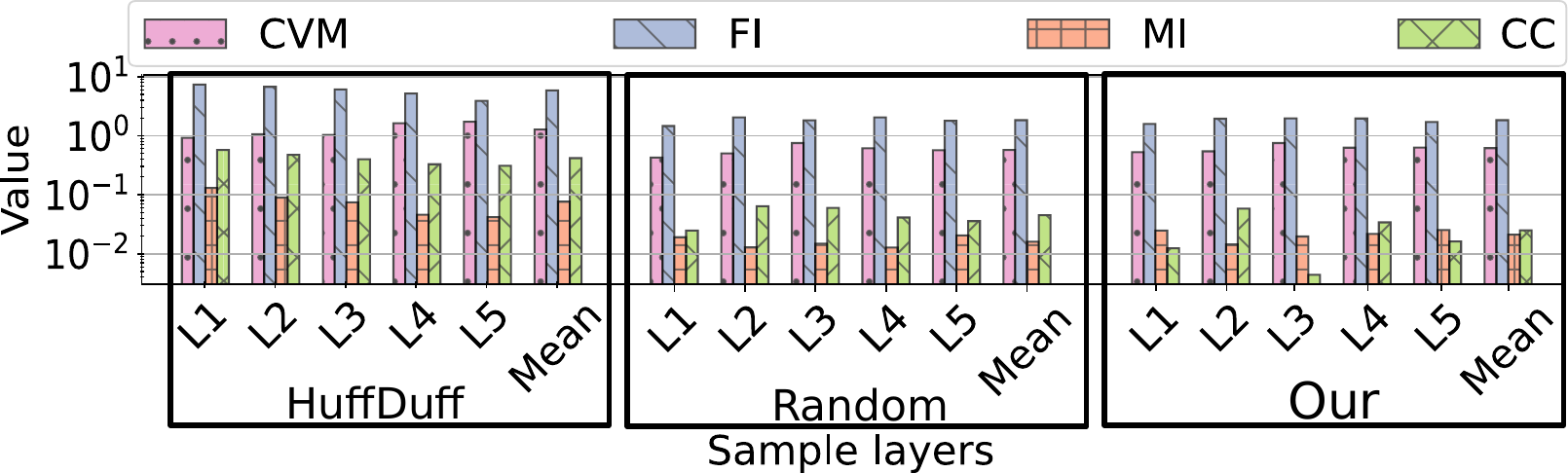}
  \caption{Security analyses for Case Study 1. \textit{MI} is b/w the secret parameter and the leaked metric. \textit{CC} is b/w the leaked traces and the filter size. }
  \label{fig:security}
\end{figure}

\textit{HuffDuff} relies on the information distortion (number of non-zero(NNZ) values) at the edges of an
\fmap due to the way the convolution operation handles
the edge pixels. This discrepancy decreases with the layer depth as shown in Figure~\ref{fig:huffduff}.
The FI in Figure~\ref{fig:security} validates
the same fact.  A
similar behavior was observed in all the other layers and NNs. Note how close
our FI is to a truly random source (within 6\%).

\noindent $\blacktriangleright$ \textit{Mutual information (MI)} We estimate the MI between the leaked traces 
and the
hidden parameter (filter size) for all the three configurations. 
We observe a similar pattern here, indicating that the amount of {\em information between the hidden 
parameter and
the leaked traces for \fname is nearly the same as the information between random data and the hidden
parameter.}

\noindent \textbf{Statistical Tests} We perform the runs test~\cite{runstest} 
(part of the NIST suite~\cite{nist}) and
CVM tests~\cite{CVM3}, which are considered as the standard approaches to quantify security.

\noindent $\blacktriangleright$ \textit{Runs test} The
average $p$-value for \fname is $0.41$, which is more (better)
than the standard threshold value of $0.05$ (corresponding to the
null hypothesis). This establishes the random nature of the source as per this test.

\noindent $\blacktriangleright$ \textit{Cramer–von Mises (CVM) test~\cite{CVM1,CVM2,CVM3}} We estimate the {\em CVM distance},
which decreases as the difference between the random data and the sampled data decreases. We observe that \fname
exhibits the lowest CVM (closest to random, see Figure\ref{fig:security}).

\noindent $\blacktriangleright$ \textit{Correlation coefficient (CC)} We calculate the classical CC to estimate the degree of correlation between the leaked signal and the hidden parameter using {\em
Pearson's CC} (see Figure\ref{fig:security}). Here also we have make a similar inference.

\subsection{Case Study 2 -- Defense from Reverse Engg.} 
We implemented ReverseEngg\cite{reverse}, which is a state-of-the-art
memory-based SCA (side-channel attack) that steals the model architecture in the absence of any CMs; it  is the
basis for many other attacks that use the same principle such as~\cite{cachetelepathy,deepsniffer}. We
implemented two CMs (based on this):
({\em NeurObfuscator~\cite{neurobfuscator}} and {\em DNNCloak}~\cite{dnncloak}) -- the former introduces dummy
computations and the latter uses a combination of dummy accesses, partial encryption and extensive on-chip buffering.
We compute the distribution for the two most important architectural hints (exposed via side channels):
RW distance and the total traffic distribution (see Table~\ref{tab:memAnalysis}). 
We observe minimal differences in the FI between \fname and {\em
DNNCloak} indicating that both the schemes are effective at mitigating address-based SCAs. However, {\em DNNCloak}
can be broken using timing and Kerckhoff-based attacks (see Section~\ref{sec:KK}), whereas \fname is immune to them.

\begin{table}[!htb]
    \centering
    \footnotesize
    \begin{tabular}{|p{0.17\linewidth}|l|l|l||l|l|l|}
    \hline
    \rowcolor{black!10}
     \textbf{Scheme} & \textbf{MI} & \textbf{CC} & \textbf{FI}& \textbf{MI} & \textbf{CC} & \textbf{FI} \\
     \cline{2-7}
        \rowcolor{black!10}
     & \multicolumn{3}{c||}{Memory traffic } & \multicolumn{3}{|c|}{RW distance } \\
     \hline
     {\em NeurObfu.} & 0.898  &  0.959   &  0.658 & 0.859  &   0.807   &  1.9   \\
     {\em DNNCloak} &  0.846 & 0.923   & 0.038  &  0.635 &  0.781  & 0.236 \\
     {\em Random} & 0.234  & 0.211   &  0.123  & 0.324  & 0.298   & 0.430 \\
     \hline
     \name   &  \textbf{0.436} & \textbf{0.243} & \textbf{0.057} &  \textbf{0.456} & \textbf{0.289 } & \textbf{0.222} \\
     \hline
    \end{tabular}
    \caption{Security analysis for Case Study 2}
    \label{tab:memAnalysis}
\end{table}




%% file: relatedWork.tex
\section{Related Work}
\label{sec:RW}

\subsection{DNN Architecture Extraction: CDTV Metrics} 
Table ~\ref{tab:attack} shows a summary of the most popular DNN architecture stealing attacks. \blue{
Hua et al.~\cite{reverse} propose a state-of-the-art attack, which builds constraint equations by observing
memory access patterns and the traffic volume. 
Similarly, DeepSniffer~\cite{deepsniffer} exploits the relationship between architectural hints
such as memory reads/write volumes and a DNN's architecture. 
The authors of HuffDuff~\cite{huffduff} exploit the {\em boundary effect} and the timing side-channel for attacking 
sparse CNN accelerators. The boundary effect occurs at the outermost edge of a convolution layer where
we perform fewer multiplications -- this
helps the attacker discern filter dimensions. 
Cache Telepathy~\cite{cachetelepathy} steals the CNN's architecture 
by estimating the timing information using  cache-based prime-probe attack. They use this information to infer the total number of 
calls for performing multiplication, which helps estimate the model architecture.  }

\begin{table}[!htb]
    \centering
    \footnotesize
\begin{tcolorbox}[enhanced, width=0.45\textwidth, boxsep=0pt, left=0pt,right=0pt,top=0pt,bottom=0pt,
    colback=white, colframe=black, arc=0mm, boxrule=0pt,
    drop shadow={shadow xshift=100mm, shadow yshift=100mm}]
 \rowcolors{2}{royalblue!20}{royalblue!5}
    \begin{tabular}{|p{24mm}|c|c|c|c|c|c|}
    \hline
    \rowcolor{gray!10}
     \textbf{Attack}  & \multicolumn{2}{c|}{\textbf{Theft target}} & \multicolumn{4}{c|}{\textbf{Leaked Metric}}  \\
       \cline{2-4}
       \cline{5-7}
      \rowcolor{gray!10}
        &  \textbf{Arch.} &  \textbf{Params} & \textbf{C} & \textbf{D} & \textbf{T} & \textbf{V} \\
    \hline
    
     ReverseEngg~\cite{reverse} &  \fullcirc[1ex] &  \fullcirc[1ex] &  \emptycirc[1ex] & \fullcirc[1ex] & \emptycirc[1ex]& \fullcirc[1ex] \\
     \hline
   
     CacheTelepathy~\cite{cachetelepathy} &  \fullcirc[1ex] &  \emptycirc[1ex] &  \fullcirc[1ex] &  \fullcirc[1ex] & \fullcirc[1ex]& \fullcirc[1ex] \\
      \hline
     
     DeepSniffer~\cite{deepsniffer} &  \fullcirc[1ex] &  \emptycirc[1ex] &   \emptycirc[1ex] &  \fullcirc[1ex] & \emptycirc[1ex]& \fullcirc[1ex] \\
    \hline

    HuffDuff~\cite{huffduff} &  \fullcirc[1ex] &  \emptycirc[1ex] &  \emptycirc[1ex] &  \fullcirc[1ex] & \fullcirc[1ex] & \fullcirc[1ex]\\
    \hline
    \end{tabular}
    \end{tcolorbox}
    \caption{Different Attacks on NNs. (C,D,T,V) stands for (Count, Distance, Time, Volume) }
    \label{tab:attack}
    \vspace{-3mm}
\end{table}

\subsection{DNN Architecture Protection}
\subsubsection{Shuffling-based CMs}  Liu et al.~\cite{mitigating} proposed to shuffle the addresses, 
while Li et al.~\cite{neurobfuscator} obscured the tile access pattern by shuffling the order of the tile accesses.  Sadly, their scheme still preserves counts and average RW distances~\cite{dnncloak}.
Additionally, shuffling requires a huge mapping table that is hard to store on chip.~\cite{dnncloak}.  

\subsubsection{Redundany-based CMs}
The number of memory accesses or the process execution time information can be hidden by incorporating dummy memory 
accesses or by adding delays.

\noindent $\blacktriangleright$ {\em Redundancy in Space} 
Li et al.~\cite{neurobfuscator} employ different obfuscation techniques such as layer deepening and 
kernel widening that introduce dummy memory accesses. This makes it necessary to
read all the dummy data at least once during the execution of the DNN, and then use or hide the results.
Che et al.~\cite{dnncloak} emphasize that attackers can distinguish between layer types by examining memory 
access intensities. 
They suggest adding dummy accesses to equalize the number of accesses across the layers, which leads
to high performance overheads. The CDTV information is nonetheless available.

\noindent $\blacktriangleright$ {\em Redundancy in Time} Wang et al. propose
{\em NPUFort}~\cite{npufort} -- a custom HW accelerator that encrypts only security-critical features.
Sadly, partial encryption of the weights only provides partial security. Similarly, 
cache-telepathy~\cite{cachetelepathy} \blue{proposes to run an entire toy NN to obfuscate 
the timing information of the DNN -- this still ends up exposing a lot of CDV information
(much more than the state of the art CMs). }

\subsubsection{Access Removal using On-Chip Buffering} 
Che et al.~\cite{dnncloak} choose to buffer a part of the \ofmap due to which an adversary will find it difficult to
infer the
size of the \ofmap (can be broken using schemes shown in Section~\ref{sec:back}). 

%% file: conclusion.tex
\section{Discussion and Conclusion}
\label{sec:conclusion}
\blue{
The key contribution of this paper is as follows. Assume a \underline{threat model} where an adversary can 
\textbf{observe} all
the addresses being sent to encrypted, tamper-proof 
DRAM memory, and \textbf{record} the volume of encrypted data transferred as well as the 
associated timing information. All known attacks and CMs on NNs \underline{use} a subset of the
CDTV metrics. We are not aware of any fifth metric that can be derived from an observation
of DRAM accesses and timing, which is fundamentally different.
In the backdrop of this observation, we create a comprehensive universal attack that relies 
on a combination of SS, KK and SI attacks to break all state-of-the-art CMs in 8-10 hours on 
current machines. The primary deficiency of current CMs is that they rely on additive noise, hardwire
parameters that can be leaked by malicious insiders and do not take cognizance of the side-information
available to an adversary. \fname is a novel CM that prevents SS, KK and SI attacks using CDTV
metrics in the context of the aforementioned threat model. Its cornerstone is the addition
of multiplicative noise of the form $Y=CX+N$, where the factor $C$ arises naturally from 
feature map compression.  This
formulation inherently leads to a mathematical framework based on Mellin transforms 
where the search space size can be
precisely quantified. Its size can further be increased by modulating $C$ (adding random
data to layers) depending on the model creator's requirements.}

\blue{
This multiplicative formulation naturally leads to a design that represents the 3D computation
in a CNN as a read-once-write-once 1D computation using SFCs (novel contribution). 
We further show that all kinds of 
corner cases including handling halo pixels and weights that do not fit in on-chip memory
can be easily handled using our SFC-based
framework. All the proposed HW structures were meticulously coded in Verilog and synthesized
for two targets: 28-nm ASIC process and the Virtex-7 FPGA. The HW overheads are quite modest 
(1.5 $mm^2$ on the ASIC).}

\blue{
Our design has a large search space size ($10^{196}$ for {\em ResNet18} and $10^{70}$ for {\em Vgg16})
with a minimal performance degradation with respect to the unsecure baseline. Moreover, for all the known
statistical tests (also part of the NIST test suite), the information leaked by
\fname is statistically very similar
to a truly random source for the CDTV metrics,
even in the presence of specially crafted inputs (Case Studies 1 and 2). With a minimal loss
in performance, it is possible to substantially increase the search space size (e.g. multiply it by $10^{50}$)
by modulating the multiplicative parameter. Owing to compression, which is intrinsically aligned with our SFC-based
framework, we note a
performance improvement of 15.48\% over our nearest competitor DNNCloak~\cite{dnncloak}.
}